\documentclass[aps,twocolumn,floatfix]{revtex4}
\usepackage{amsmath}
\usepackage{latexsym}
\usepackage{float}
\usepackage{amssymb}
\usepackage{graphicx}
\usepackage{epsfig}
\usepackage{color}

\definecolor{red}{rgb}{0.0,0.0,0.0}

\begin{document}

\title{Lattice Monte Carlo Simulations of Polymer Melts}

\author{Hsiao-Ping Hsu} 
\affiliation{Max-Planck-Institut f\"ur Polymerforschung, Ackermannweg 10, D-55128 Mainz, Germany}
\begin{abstract}
   We use Monte Carlo simulations to study polymer melts consisting of
fully flexible
and moderately stiff chains in the bond fluctuation model at
a volume fraction $0.5$. In order to reduce the local density
fluctuations, we test a pre-packing process for the preparation of the
initial configurations of the polymer melts, before
the excluded volume interaction is switched on completely.
This process leads to a significantly faster decrease of the number of
overlapping monomers on the lattice. 
This is useful for simulating very large systems, where
the statistical properties of the model with a marginally incomplete
elimination of excluded volume violations are the same as those of the
model with strictly excluded volume.
We find that the internal mean square end-to-end distance for moderately stiff
chains in a melt can be very well described by a freely rotating chain
model with
a precise estimate of the bond-bond orientational correlation between
two successive
bond vectors in equilibrium.
The plot of the probability distributions of the reduced end-to-end
distance of chains of
different stiffness also shows that the data collapse is excellent and
described
very well by the Gaussian distribution for ideal chains.
However, while our results confirm
the systematic deviations between Gaussian statistics for the
chain structure factor $S_c(q)$ [minimum in the Kratky-plot] found by
Wittmer et al.~\{EPL {\bf 77} 56003 (2007).\} for fully flexible
chains in a melt, we show that for the available chain length these
deviations are no longer visible, when the chain stiffness is included.
The mean square bond length and the compressibility estimated from
collective structure factors
depend slightly on the stiffness of the chains.
\end{abstract}

\maketitle

\section{Introduction}

In the theoretical study of polymer physics~\cite{Flory1969,deGennes1979},
computer simulations provide a powerful method
to mimic the behavior of polymers covering the range from atomic to coarse-grained scales
depending on the problems one is interested 
in~\cite{Binder1995,Kotelyanskii2004,Binder2008}.
The generic scaling properties of single linear and branched polymers in the bulk
under various solvent conditions and polymer solutions at different temperatures
have been described quite well by simple coarse-grained lattice models, e.g. 
self-avoiding walks on the simple cubic lattice, and off-lattice models, e.g.
bead-spring model, etc. 
A wide variety of computational strategies has been employed to simulate 
and analyze these models, including conventional (Metropolis) Monte Carlo (MC)
schemes with various types of moves, chain growth algorithms, parallel
tempering, and molecular dynamics~\cite{Binder1995}. 
On the one hand, however, as the size and complexity of a system increases, detailed 
information at the atomic scale may
be lost when employing low resolution coarse-graining representations. 
On the other hand,
the cost of computing time may be too high if the system is described at high resolution.
Therefore, more scientific effort has been devoted to developing
an appropriate coarse-grained model where the connection to the details
on the atomic scale is not lost, but rather atomistic details are suitably mapped on
effective potentials on the coarser scale. Such models then
can reproduce the global thermodynamic properties and the local mechanical and chemical properties
such as the intermolecular forces between polymer
chains~\cite{Murat1998, Plathe2002, Harman2006,Harman2007, Gujrati2010, Vettorel2010,
Zhang2013, Zhang2014}.
Improving such approach further is still an active area of research.

In this work we deal with linear polymer chains in a melt on the simple cubic lattice. 
Although coarse-grained lattice models neglect the chemical detail of a specific
polymer chain and only keep chain connectivity (topology) and excluded volume,
the universal behavior of polymers still remains the same in the thermodynamic
limit (as the chain length $N\rightarrow \infty$)\cite{deGennes1979}. 
We consider for our simulations the bond fluctuation model 
(BFM)~\cite{Carmesin1988, Wittmann1990, Deutsch1991, Paul1991, Binder1995} 
where linear chains are located on the simple cubic lattice with bond constraints.
The BFM has the advantages that the computational efficiency of lattice models is
kept and the behavior of polymers in a continuum space can be described
approximately. The model thus introduces some local conformational flexibility while
retaining the computational efficiency of lattice models for implementing
excluded volume interactions by enforcing a single occupation of each lattice vertex.
Although much work using this model exists already, only the fully
flexible limit of the model has been used exclusively. This limit does not suffice when
one considers a possible mapping of atomistic details to this model, which requires
to include some description of chain stiffness as well.

A review of recent BFM studies is given in Ref.~\cite{Wittmer2011}.
When the concentration of polymer solutions is above the
overlap concentration denoted by $c^*$, the excluded volume
interactions are screened~\cite{deGennes1979,Yamakawa1971}.
The average interaction between monomers finally should cancel
in a polymer melt since every monomer is isotropically surrounded
by other monomers belonging to the same chain or not according to Flory's
argument. Therefore the polymer chains in a melt behave as ideal chains,
where the excluded volume effect is no longer important.
However, a careful investigation of an individual polymer chain in
a polymer melt based on the BFM and the bead-spring model (BSM) shows 
that there are noticeable 
deviations from an ideal chain behavior~\cite{Wittmer2007i,Wittmer2011}. 
{The deviations are due to the incompressibility
constraint of the melt. Therefore, small scale-free corrections
to the asymptotic behavior for ideal chains exist and 
they are irrespective of the stiffness of the chains.
In Refs.~\cite{Wittmer2011,Wittmer2007i}, the authors
have shown that the corrections must decay as the
dimensionless (Ginzburg) parameter 
$\rho^*/\rho \sim 1/(\rho b_e^3 s^{1/2})$ for 
$1 \ll s \ll N$ with
the increase of the segments $s$ and the effective
bond length $b_e$ of the chain.
Here $\rho$ is the overall monomer density and $\rho^* \sim s/R(s)^3$
the segmental overlap density related to the internal distance
$R(s) \approx b_e s^{1/2}$.
It is still unclear whether the predicted power-law deviations
from unperturbed Gaussian behavior can be observed for the internal segment
lengths of semiflexible chains
in the range $1 \ll g(\varepsilon_b) \ll s \ll N$ by varying the 
stiffness of chains, i.e. tuning the bending energy $\varepsilon_b$.}

A very important practical problem for the simulation of melts
of very long polymer chains is the construction of a suitable initial
configuration.
A naive approach for creating initial configurations of melts with long
polymers, would be to switch off initially the excluded volume interactions
between monomers (so that the polymers resemble Gaussian chains), arrange
these chains randomly in a simulation box, and switch on 
gradually the excluded volume
interactions at the final step.
It has been demonstrated by simulations using the BSM
in the continuum that this strategy is not
feasible since it leads to chain deformations on short length
scales~\cite{Auhl2003}.
With increasing chain length this effect becomes more pronounced due to the fact
that self-screening and correlation hole effects, inherent to flexible high
polymer melts, are not properly accounted for by this procedure.
This problem was overcome by reducing the density fluctuations of
Gaussian chains via a ``pre-packing" strategy, and then switching on the
excluded volume interactions in a quasi-static way
(“slow push-off”)~\cite{Auhl2003,Moreira2014}. 
We test a similar strategy of simulating
polymer melts based on the lattice model BFM in this work to check whether it
helps to generate nearly equilibrated initial configurations through such 
a procedure.

The outline of the paper is as follows: Sec.~\ref{model}
describes the model and the simulation technique discussing how
to generate initial configurations of polymer chains in a melt
to equilibrate the system.
Sec.~\ref{results} presents the simulation results at
different stages from the initial state to the equilibrium state
of fully flexible and moderately stiff polymer chains in a melt. 
Finally our conclusions are summarized in Sec.~\ref{conclusion}.

\section{Model and simulation techniques}
\label{model}
In the standard bond fluctuation model
(BFM)~\cite{Carmesin1988, Wittmann1990, Deutsch1991, Paul1991, Binder1995}
a flexible polymer chain with excluded volume interactions is 
described by a self-avoiding walk (SAW) chain of effective monomers on 
a simple cubic lattice (the lattice spacing is the unit of length).
Each effective monomer of such a SAW chain blocks all
eight corners of an elementary cube of the simple cubic lattice
from further occupation. Two successive monomers along a chain are 
connected by a bond vector $\vec{b}$ which is taken from the set
\{$(\pm2,0,0)$,$(\pm 2,\pm 1,0)$,
$(\pm 2, \pm 1, \pm 1)$, $(\pm 2, \pm 2, \pm 1)$, $(\pm 3, 0,0)$,
$(\pm 3,\pm 1,0)$\} including also all permutations. The bond length
$\mid \vec{b} \mid$ is therefore in a range between $2$ and $\sqrt{10}$.
There are in total
$108$ bond vectors and $87$ different bond angles between two sequential bonds
along a chain serving as candidates for building the conformational
structure of polymers.

 As the stiffness of the chains is considered, semiflexible chains of $N$ monomers in a melt
thus are described by SAW chains on the simple cubic lattice, with a bending
potential
\begin{eqnarray}
    \frac{U_b}{k_BT}&=&\frac{\varepsilon_b}{k_BT} \sum_{i=1}^{N-1}(1-\cos \theta_{i,i+1}) \nonumber \\
& =& \frac{\varepsilon_b}{k_BT}\sum_{i=1}^{N-1}\left(1-\frac{\vec{b}_i \cdot \vec{b}_{i+1}}
{\mid \vec{b}_i \mid \mid \vec{b}_{i+1} \mid} \right) \,,
\end{eqnarray}
where $\varepsilon_b$ is the bending energy ($\varepsilon_b=0$ for ordinary SAWs),
and $\theta_{i,i+1}$ is the bond angle between
the $i^{\rm th}$ bond vector and the $(i+1)^{\rm th}$ bond vector along a chain.
$k_BT$ is of order unity throughout the whole paper.
A systematical study of single semiflexible chains based on the
BFM~\cite{Hsu2014} shows that due to bond vector fluctuations and lattice artifacts
the initial decay of the bond-bond orientational correlation
function deviates from the simple
exponential decay for ${\varepsilon_b>10}$. Therefore, one should be careful of using
the BFM for studying rather stiff chains as has already been pointed out
also in Ref.~\cite{Wittmer1992}.

It is well understood that linear polymers in a melt can be described
by SAW chains based on the BFM at a volume fraction 
$\phi=0.5$~\cite{Deutsch1991, Paul1991, Binder1995}.
Since in the BFM, each effective monomer occupies one unit cell,
containing $8$ lattice sites, the monomer density is therefore defined
as $\rho=\phi/8$.
A detailed description of how we prepare the initial configurations is given as follows:
The initial configuration of polymer melts containing $n_c$ chains of $N$ monomers
in a box of size $V=L^3$ with periodic boundary conditions
in all three directions are generated in the following way:
\begin{itemize}
\item The linear dimension of a simple cubic lattice is set up as
$L=(8n_cN/\phi)^{1/3}$ with volume fraction $\phi=0.5$.
\item At the $0^{\rm th}$ step the first $n_c$ monomers of the chains are
randomly put on the lattice sites without double overlapping.
It can be easily done by dividing the box into $n_c$ blocks
and putting a single monomer in each block at a randomly chosen
position inside the block.
\item Polymer chains are built like non-reversal random walks (NRRWs) by adding
one monomer at each step until all $n_c$ chains reaching the required chain
length $N-1$. (We define here the chain length as the number
of bonds along the contour of a chain. The contour length of the chain  
then is $L=(N-1)\ell_b$ where $\ell_b=\langle \mid {\vec{b}}^2 \mid \rangle^{1/2}$ 
is the root-mean-square bond length).
Based on the BFM there are $108$ possibilities to place the next
monomer for each chain at the $1^{\rm st}$ step, but at the following steps only
one of $107$ directions can be selected since an immediately reverse step is
not allowed.  At this stage, the excluded volume effect is switched off completely
and $n_c$ NRRWs of $(N-1)$-steps in the simulation box
are generated. 

If the stiffness of the chains is considered,
the probability of placing the $(i+1)^{\rm th}$ monomer connected to the $i^{\rm th}$ monomer
is proportional to $\exp(-\varepsilon_b(1-\cos \theta_{i,i+1}))$.

\item Finally the excluded volume interactions between monomers are considered
by applying the Local 26 and slithering-snake moves to relax
the polymer chains and push off those monomers blocking the same lattice site
until all chains satisfy self- and mutual-avoidance. This ``steepest descent"
approach only works for lattice chains since excluded volume constraints
can be checked very efficiently, but would not work for continuum chains.
\end{itemize}

Due to the large local density fluctuations of randomly overlapping
NRRW chains in a melt, a pre-packing procedure
to reduce the chain deformations is also tested before 
the excluded volume interactions are switched on.
This pre-packing procedure
has been successfully performed in the simulations of equilibrating
very long polymer chains in melts up to $n_c=1000$ and $N=1000$
based on the bead-spring model in the continuum~\cite{Moreira2014}.
In this procedure all NRRW chains still keep their own structures as
rigid bodies, but are rearranged by Monte Carlo (MC) trial moves.
The cost function which describes the average fluctuations of the particle
number $n(\sigma)$
within a sphere of radius $\sigma$ with its center located at any
monomer $i$ is defined by
\begin{eqnarray}
    U(\sigma) &=& <n(\sigma)^2>-<n(\sigma)>^2 \\ \nonumber
&=&\frac{1}{N_{\rm tot}} \sum_{i=1}^{N_{\rm tot}} n_i^2(\sigma) 
- \left(\frac{1}{N_{\rm tot}} \sum_{i=1}^{N_{\rm tot}} n_i(\sigma) \right)^2
\label{eq-fluc}
\end{eqnarray}
with
\begin{equation}
  n_i(\sigma)=\sum_{j=1\;,j \neq i}^{N_{\rm tot}} H(\sigma-r_{ij})
\,, \quad H(x)=\left\{\begin{array}{ll} 
        0  \,, & {\rm for} \; x<0\\
        1\,, & {\rm for} \; x\geq 0
\end{array} \right .
\end{equation}
where $N_{\rm tot}=Nn_c$ is the total number of monomers in a melt,
$n_i(\sigma)$ is the number of monomers inside the $i^{\rm th}$
sphere of radius $\sigma$, $r_{ij}=\mid \vec{r}_i - \vec{r}_j \mid$ is
the distance between monomers $i$ and $j$ irrespective of
the chain connectivity, and $H(x)$ is the Heaviside step
function. A trial move is accepted if the cost function becomes smaller,
namely, the local density fluctuation is minimized in this process.
Two types of MC moves, translation and pivot-like moves,
applied to the center of mass of an individual chain are considered
as follows:
(a) In the translation move, an individual chain (or its center
of mass) is randomly translated by a vector chosen from the set
$\left\{(\pm 1,0,0),(\pm 1, \pm 1,0),(\pm 1,\pm 1,\pm 1)\right\}$
including all permutations.
(b) In the pivot-like move,  a whole individual chain is transformed by
randomly choosing one of the $47$ symmetric operators including
rotations by $90^o$ or $180^o$ around a random axis and reflections
through its center of mass, and inversion at its center of mass.
The center of mass serves as a pivot point here.
Note that all monomers are only allowed to sit on the lattice sites,
so a chain would need to be shifted slightly
after the pivot-like move. However, the deviation of its center of mass
is less than $1$ lattice spacing.

For equilibrating the system and measuring interesting physical
quantities in the equilibrium runs
three types of MC moves, local $26$ moves~\cite{Wittmer2007}, 
slithering-snake moves,
and pivot moves, are used in our simulations and briefly described
as follows,
\begin{itemize}
\item Local $26$ (``L26") move: a monomer is chosen randomly to
move to one of the $26$ nearest and next nearest neighbor sites
surrounding it, i.e, randomly
translated by a vector chosen from the set
$\left\{(\pm 1,0,0),(\pm 1, \pm 1,0),(\pm 1,\pm 1,\pm 1)\right\}$
including all permutations. The bond crossing is allowed during the move.
This is different from the traditional local $6$ (``L6") move where
a monomer of the chain is chosen at random and moved to the
nearest neighbor sites in the six lattice directions randomly. 
\item Slithering-snake move: an end monomer is removed randomly and 
connected to the other end of the same linear chain by a bond
vector randomly chosen from the set of $108$ allowed bond vectors.
\item Pivot move: a monomer is chosen randomly from a linear chain
as a pivot point, and the short part of the linear chain is transformed
by randomly selecting one of the $48$ symmetry operations (no change;
rotations by $90^o$ and $180^o$; reflections and inversions).
\end{itemize}
Of course, any attempted Monte Carlo move is accepted only if it does
not violate constraints of our physical systems, such as
excluded volume and bond length constraints.

\section{Simulation results}
\label{results}

  We first wish to test whether applying the pre-packing process to rearrange the
NRRW chains based on the BFM before the excluded volume 
interaction is switched on helps to prepare the nearly equilibrated initial 
configurations or not.
We restrict the size of the lattice to be $L^3=128^3$ in our simulations.
The total number of effective monomers
based on the BFM is therefore $N_{\rm tot}=n_cN=131072$.
Varying $n_c$ and $N$ but keeping $N_{\rm tot}$ fixed the conformational
properties of polymer melts including fully flexible chains ($\varepsilon_b=0.0$)
and moderately stiff chains ($\varepsilon_b=1.0$, $2.0$, and $3.0$) are studied. 
Results of simulating polymer melts from preparing the initial configuration
to the analysis of equilibrated chain structures in a melt are discussed in 
the next Subsection III A.

\subsection{Local density fluctuations}

\begin{figure}[t]
\begin{center}
\includegraphics[scale=0.29,angle=270]{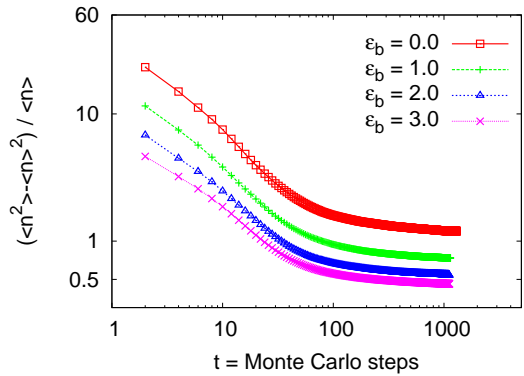}
\caption{Time series of local density fluctuation
$(\langle n(\sigma)^2 \rangle - \langle n(\sigma) \rangle^2)/\langle n(\sigma) \rangle$
for $\sigma=12$. Data are for $n_c=256$ NRRW chains
of $N=512$ monomers in a box of size $L^3=128^3$. Several bending energies
$\varepsilon_b$ are chosen, as indicated.}
\label{fig-melt-u}
\end{center}
\end{figure}

In Fig.~\ref{fig-melt-u} we show the typical time series of the
local density fluctuation
$(\langle n(\sigma)^2 \rangle - \langle n(\sigma) \rangle^2)/\langle n(\sigma) \rangle$ with $\sigma=12$
for $n_c=256$ NRRW chains of $N=512$ monomers in a box of size $L^3=128^3$, and
for the bending energies $\varepsilon_b=0$ (fully flexible), $1.0$, $2.0$, and $3.0$.
In the pre-packing process, the radius $\sigma$ is first set to $12 \approx 4 \ell_b$ in the unit
of lattice spacings until all curves describing the local density fluctuation reach a plateau, and 
then reduced to $6 \approx 2 \ell_b$ until all curves reach another plateau again.
The estimates of 
the collective structure factor $S(q)$ for $q=2\pi/L$ and the 
local density fluctuation
$(\langle n(\sigma)^2 \rangle - \langle n(\sigma) \rangle^2)/\langle n(\sigma) \rangle$ for 
$\sigma=12$ are of the same order of magnitude (see Sec.~\ref{structure}) from the final
configurations generated by this two-step pre-packing process.
Each Monte Carlo step contains $n_c$ translation moves and $n_c$ pivot-like moves
such that the locations of all chains in the box are rearranged by the trial moves.
Each curve shown in Fig.~\ref{fig-melt-u} represents
the results averaged over $32$ independent realizations.
We see that the local density fluctuation can be reduced by about an order of
magnitude after about $1000$ Monte Carlo steps.

\begin{figure*}[t]
\begin{center}
(a)\includegraphics[scale=0.29,angle=270]{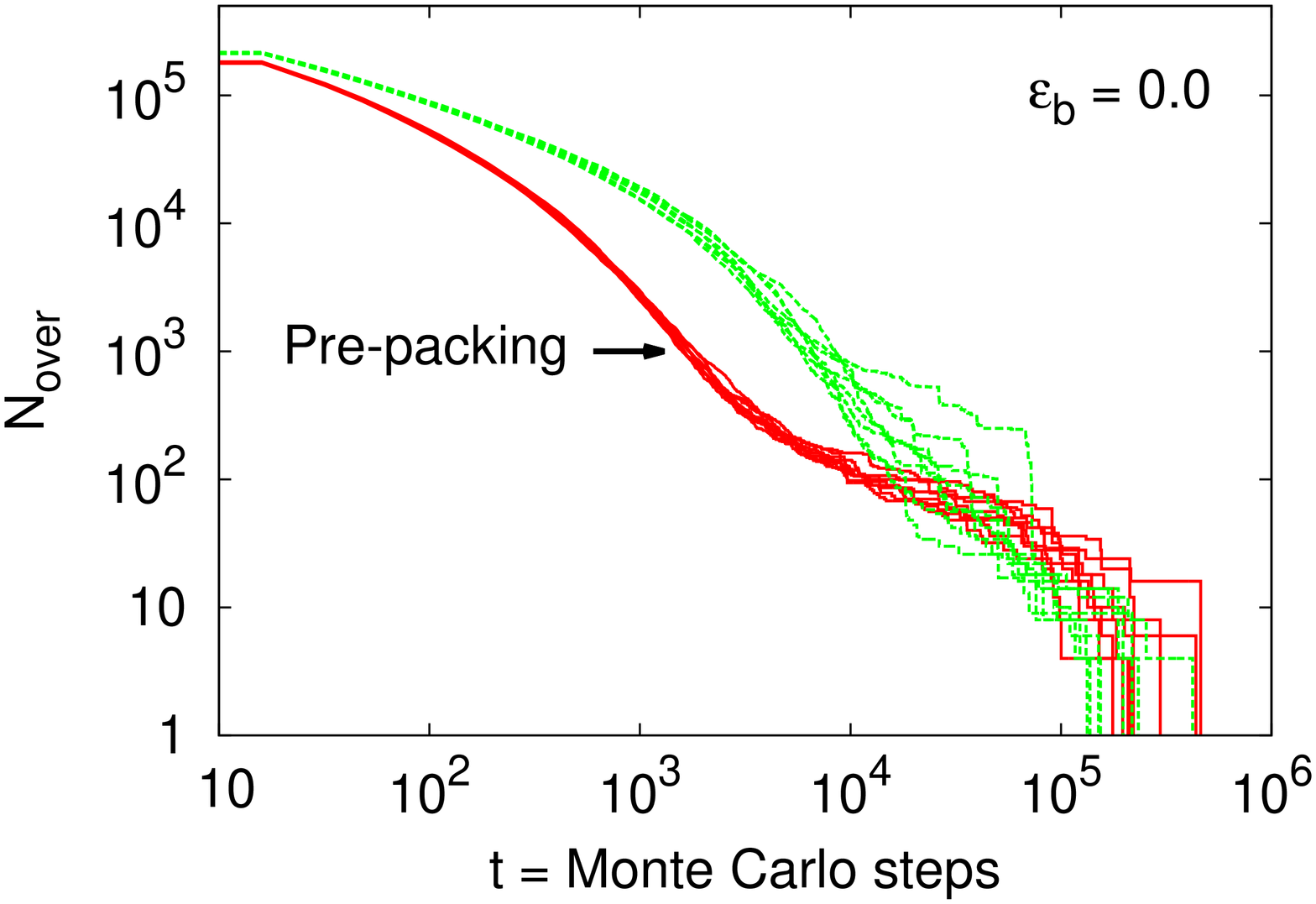} \hspace{0.4cm}
(b)\includegraphics[scale=0.29,angle=270]{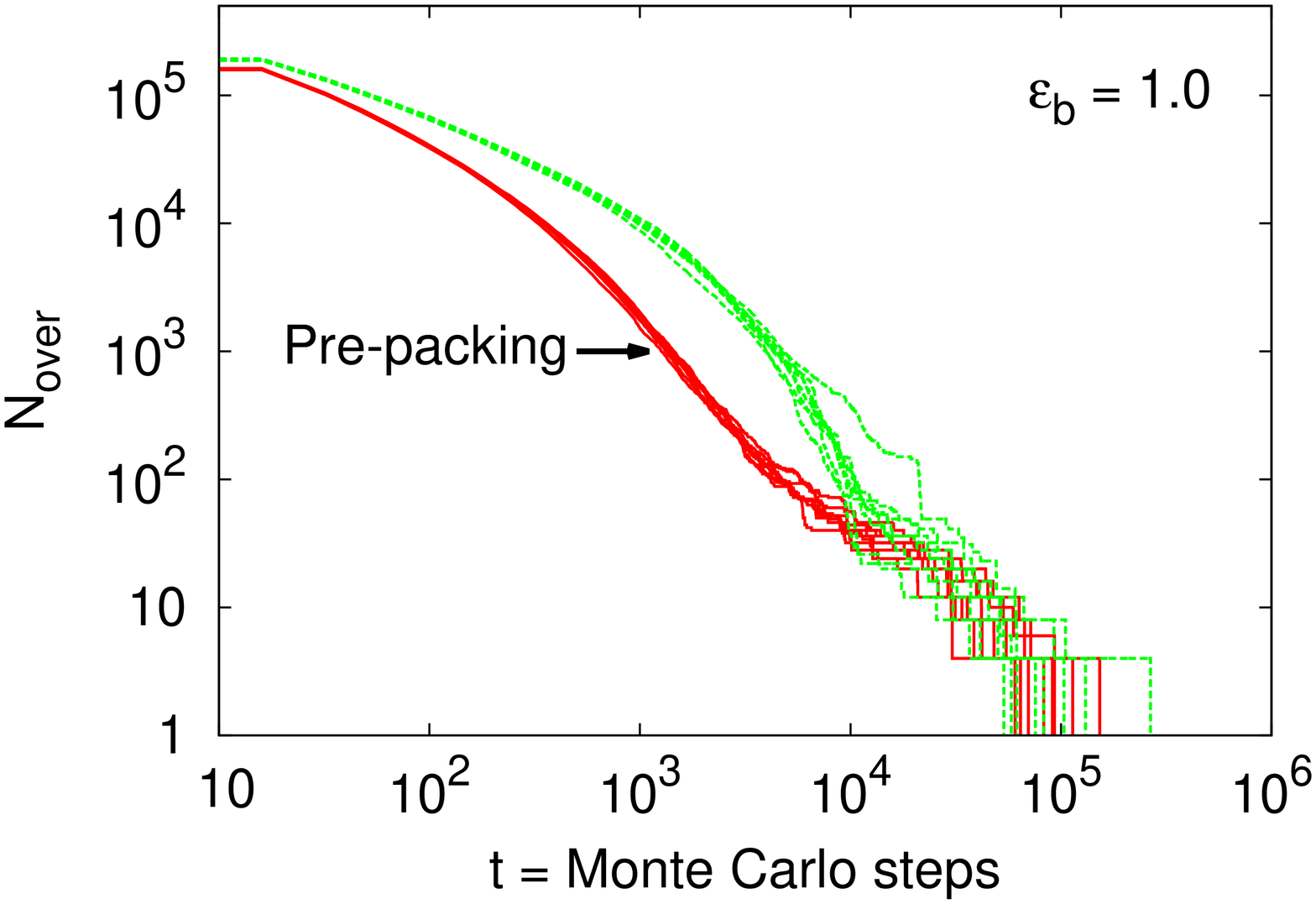}\\
(c)\includegraphics[scale=0.29,angle=270]{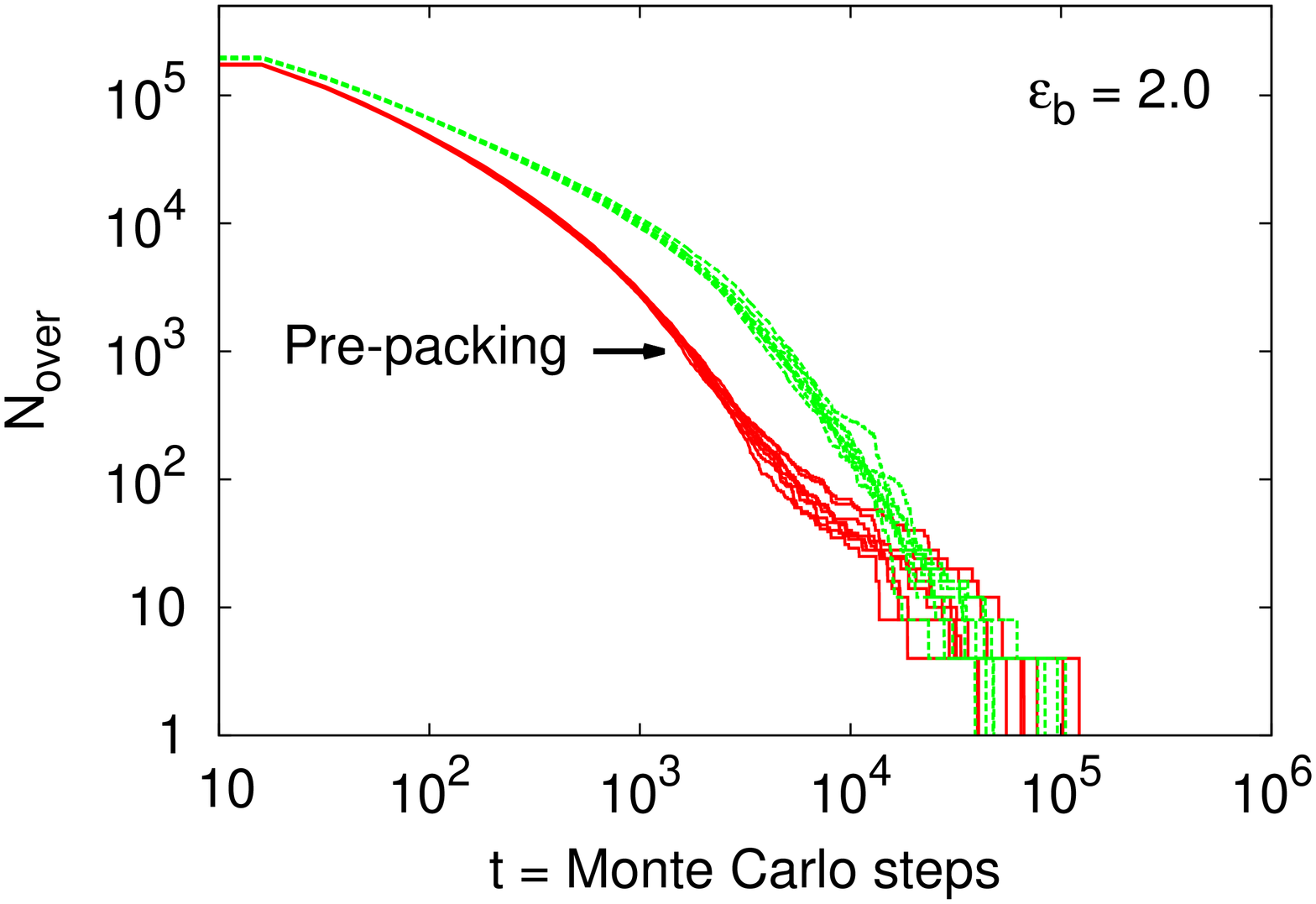} \hspace{0.4cm}
(d)\includegraphics[scale=0.29,angle=270]{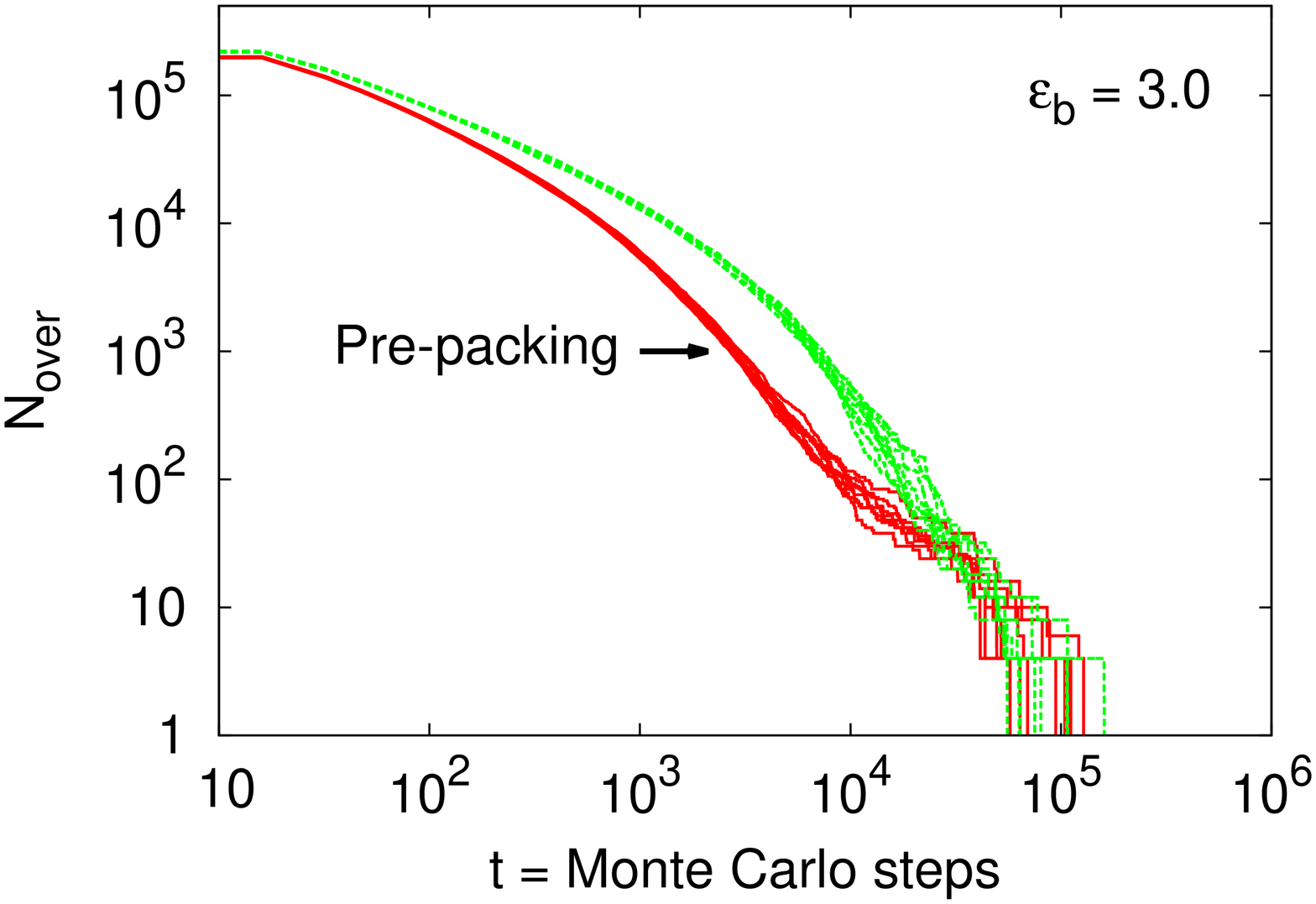}\\
\caption{Time series of the total number of 
overlapping monomers $N_{\rm over}$ 
for polymer melts consisting of $n_c=256$ chains of $N=512$ monomers, 
and for $\varepsilon_b=0.0$ (a), $1.0$ (b), $2.0$ (c), and $3.0$ (d). 
Two sets of data for the simulations with (solid curves in red color online)
and without (dashed curves in green color online) the pre-packing 
process, as indicated.}
\label{fig-melt-overlap}
\end{center}
\end{figure*}

It is interesting to know what we have gained by introducing
the pre-packing procedure before equilibrating the system. 
Therefore, we check the process of
pushing away the monomers occupying the same unit cells (eight lattice sites
are occupied by one effective monomer in BFM) by observing the time series of
the number 
of overlapping monomers, $N_{\rm over}$, as the excluded volume effect 
is switched on. 
For simplicity, if a lattice site is occupied by $s$ effective monomers, it 
contributes $s-1$ overlapping monomers to $N_{\rm over}$. 
The transition process from $n_c=256$ NRRW chains of $N=512$ monomers
to SAW chains is observed according to the decrease of $N_{\rm over}$ 
shown in Fig.~\ref{fig-melt-overlap} for $\varepsilon_b=0.0$,
$1.0$, $2.0$, and $3.0$. 
Two sets of data ($8$ samples for each) representing 
the push-off process before (dashed curve in green color online) and after 
(solid curves in red color online) the pre-packing process is applied. 
Here one MC step is a sequence 
of $N_{\rm tot}$ L26 moves, and $Nn_c$ slithering-snake moves.
In the slithering-snake moves, 
an end-monomer of each chain is selected and attempted to
move once until all chains have been chosen, and then the 
same procedure is repeated $N$ times. 
A trial move is accepted if 
$\Delta N_{\rm over} =N_{\rm over}^{\rm (new)}-N_{\rm over}^{\rm (old)} < 0$
and $\exp(-\Delta U_b)/k_BT>r$ where $r$ is a random number and $r \in [0,1)$.
The latter condition is the Metropolis criterion preserving 
the stiffness of the chains.
If only L26 moves are applied in this push-off process, at the beginning 
it is very efficient to push off overlapping monomers, but then some chains
are trapped in a state containing knots, or 
a state blocked by other chains.
The slithering-snake moves therefore indeed play an important role
at the intermediate stages for the chains escaping from a trapped state.
From our observations in Fig.~\ref{fig-melt-overlap} it seems that applying an 
additional pre-packing process does not speed up the push-off process.
However, it is still necessary to investigate the conformation of chains
in a melt carefully.

In addition, it is interesting to note that pre-packing leads to a
significantly faster decrease of $N_{\rm over}$ with time during intermediate
stages of the process. This feature may be of interest in cases where one
does not require that strictly $N_{\rm over}=0$ in the initial stage of
the averaging. Consider e.g. a variant of the model where monomers overlap is
not strictly forbidden (which corresponds to an infinitely high repulsive
energy) but only leads to a large but finite energy penalty, e.g.
$E_{\rm over}=20 k_BT$. The probability to accept a move which 
increases the overlap then is of order $10^{-9}$, i.e. essentially negligible,
and for long chains the statistical properties of the model are essentially
the same as for the model with strict excluded volume. For such a model
the overlap energy per monomer reduces after about $10^4$ Monte Carlo steps
from an initial value (of the order of $20 k_BT$) by three orders of magnitude, 
i.e. $0.02k_BT$, and this small number is reached $3$ to $4$ times faster
with pre-packing than without it. For very large systems (and complicated
models, leading to slow simulation algorithms) it may hence be an acceptable 
compromise to allow a small fraction ($10^{-3}$ or less) of overlap
in the system.

In Table~\ref{table1} we compare the average values of local density fluctuation 
$(\langle n(\sigma)^2 \rangle - \langle n(\sigma) \rangle^2)/\langle n(\sigma) \rangle$ 
\{Eq.~(\ref{eq-fluc})\} for $\sigma=12$ (since $q_{\rm min} =2\pi/L \approx V_{\rm sub}^{-1/3}$)
over $32$ configures at different stages 
from the preparation of the initial configurations to
the final configurations at the equilibrium state. 
Two ways of preparing the initial configurations (Sec.~\ref{model})
are compared as follows:
(1) Randomly generating weakly-interactive NRRWs (NRRWs), applying the pre-packing process
(Pre-packing, NRRWs), and switching on the excluded volume effect 
(Pre-packing+EV, SAWs). (2) Randomly generating weakly-interactive NRRWs (NRRWs) and
switching on the excluded volume effect (EV, SAWs). We see that for weakly-interactive
NRRWs the local density
fluctuation decreases as the bending energy $\varepsilon_b$ increases. Since
the bond angles tend to become smaller, the distance between non-bonded 
monomers apart from each other along the chain becomes longer. 
After the pre-packing process the fluctuation reduces by about
$95\%$. Once the excluded volume interactions between monomers 
are switched on,
the fluctuation remains almost the same for all cases.

\begin{table*}[ht]
\caption{Values of local density fluctuation 
$(\langle n^2 \rangle -\langle n \rangle^2)/\langle n \rangle$ for $\sigma=12$.
Data are for $N=512$, and for $\varepsilon_b=0.0$, $1.0$, $2.0$, $3.0$, averaged
over 32 configurations at different stages.}
\label{table1}
\begin{center}
\begin{tabular}{|c|rrrr|}
\hline
$\varepsilon_b$ & 0.0 & 1.0 & 2.0 & 3.0 \\
\hline
NRRWs                 & 51.06 & 30.96 & 20.89 & 14.57 \\
Pre-Packing, NRRWs   &  2.92 &  1.56 &  1.04 &  0.86 \\
Pre-Packing+EV, SAWs &  0.28 &  0.28 &  0.28 &  0.28 \\
EV, SAWs            &  0.27  &  0.27 &  0.28 &  0.27 \\
Equilibrium, SAWs   &  0.28  &  0.28 &  0.27 &  0.27 \\ 
\hline
\end{tabular}
\end{center}
\end{table*}

\subsection{Mean square internal end-to-end distance and bond-bond orientational
correlation function}

For understanding the connectivity
and correlation between monomers
the conformations of linear chains of contour length $L=(N-1)\ell_b$ in a melt 
are normally described
by the average mean square internal end-to-end distance, 
$\langle R^2(s) \rangle$,
\begin{equation}
    \langle R^2(s) \rangle = \left \langle \frac{1}{n_c}
\sum_{n=1}^{n_c} \left[ \frac{1}{N-s} \sum_{j=1}^{N-s} (\vec{r}_{n,j}-\vec{r}_{n,j+s})^2 
\right] \right \rangle \, ,
\end{equation}
where $s$ is the chemical distance between the $j^{\rm th}$ monomer and
the $(j+s)^{\rm th}$ monomer along the identical chain.
The theoretical prediction of the internal mean square end-to-end distance
for polymer melts consisting of semiflexible chains in
the absence of excluded volume effect described by a freely rotating chain model 
is~\cite{Flory1969}
\begin{equation}
   \langle R^2(s) \rangle =s\ell_b^2 \left[ 
\frac{1+\langle \cos \theta \rangle}{1- \langle \cos \theta \rangle}
-\frac{2 \langle\cos \theta \rangle( 1-(\langle \cos \theta \rangle)^s}
{s(1-\langle \cos \theta \rangle)^2} \right]  \, ,
\label{eq-Rs2-FRC}
\end{equation}
with
\begin{equation}
     \ell_b^2=\langle \vec{b}^2 \rangle \qquad {\rm and} \qquad 
 \langle \cos \theta \rangle = \langle \vec{b}_i \cdot \vec{b}_{i+1} \rangle /\ell_b^2
\end{equation}
where $\ell_b$ is the root-mean-square bond length.
In the limit $N\rightarrow \infty$,
the bond-bond orientational correlation function 
in the absence of excluded volume effects therefore 
decays exponentially as a function of 
chemical distance $s$ between any two bonds along a linear 
chain~\cite{Grosberg1994,Rubinstein2003},
\begin{equation}
 \langle  \vec{b}_i \cdot \vec{b}_{i+s} \rangle =\ell_b^2 
\langle \cos \theta(s) \rangle =\ell_b^2 \langle \cos \theta \rangle^s
=\ell_b^2 \exp(-s \ell_b / \ell_p) \,,
\label{eq-cos-dWLC1}
\end{equation}
where $\ell_p$ is the so-called persistence length which can be
extracted from the initial decay of $\langle \cos \theta (s) \rangle$.
Equivalently, one can calculate the persistence length from
\begin{equation}
    \ell_{p,\theta}/\ell_b = -1/\ln (\langle \cos \theta \rangle) 
\label{eq-lp-dWLC}
\end{equation}
here instead of $\ell_p$ we use $\ell_{p,\theta}$ to distinguish between
these two measurements. Replacing $s$ by $N-1$ in Eq.~(\ref{eq-Rs2-FRC})
it gives the asymptotic behavior of the mean square end-to-end distance of
a FRC equivalent to the behavior of a freely jointed chain 
\begin{eqnarray}
  \langle R_e^2 \rangle &=& C_\infty (N-1)\ell_b^2  \quad {\rm with}
\quad  C_\infty=\frac{1+\langle \cos \theta \rangle}{1-\langle \cos \theta \rangle}
\label{eq-Re2-N} \\
&=&n_K\ell_K^2= 2 \ell_p L
\label{eq-Re2-FRC}
\end{eqnarray}
where $C_\infty$ is called Flory's characteristic ratio~\cite{Flory1969},
$\ell_K=2\ell_p$ is the Kuhn length, and $n_K$ is the number of Kuhn segments.

\begin{figure}[t]
\begin{center}
\includegraphics[scale=0.29,angle=270]{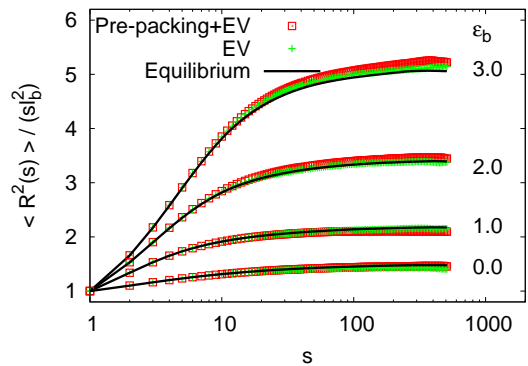} \hspace{0.4cm}
\caption{Rescaled internal mean square end-to-end distance,
$\langle R^2(s) \rangle/(s \ell_b^2)$, plotted as a function
of $s$ for polymer melts containing $n_c=256$ chains of $N=512$ monomers,
and for $\varepsilon_b=0.0$, $1.0$, $2.0$, and $3.0$. Data are
for the estimates obtained from the initial configurations of
SAWs generated by the two methods, "Pre-packing+EV", and "EV", and
from the equilibrated configurations.}
\label{fig-rs-melt512}
\end{center}
\end{figure}

We compare the estimates of $\langle R^2(s) \rangle$ obtained from the 
initial configurations of SAW chains
generated by pushing off monomers occupying the same unit cell
before and after the pre-packing process to that from those configurations
in equilibrium as shown in Fig.~\ref{fig-rs-melt512}. The two ways of preparing the initial
configurations with and without the pre-packing process are denoted by 
``EV" and ``Pre-packing+EV", respectively in the figure. Results are obtained by
taking the average over $32$ initial configurations of SAW chains for each case.
We see that there is only a slight discrepancy in the estimates of $\langle R^2(s) \rangle$ 
obtained from the initial configurations and from the equilibrated configurations.
The discrepancy becomes more prominent for $\varepsilon_b=3.0$.
We see that there is no significant advantage of preparing the initial configurations of lattice
chains in a melt through the pre-packing process as that was observed for the continuum chains.

\begin{figure*}[t]
\begin{center}
(a)\includegraphics[scale=0.29,angle=270]{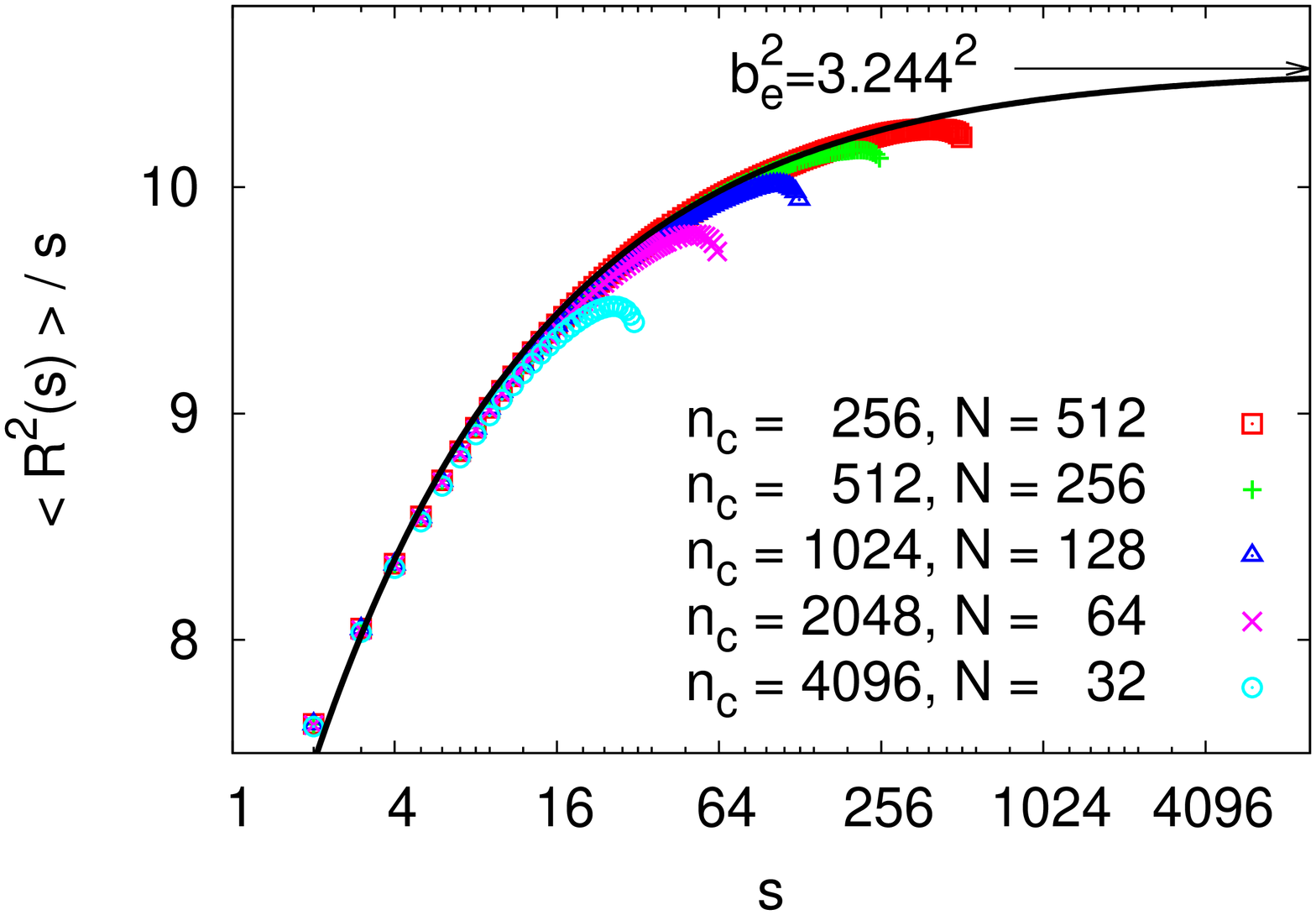} \hspace{0.4cm}
(b)\includegraphics[scale=0.29,angle=270]{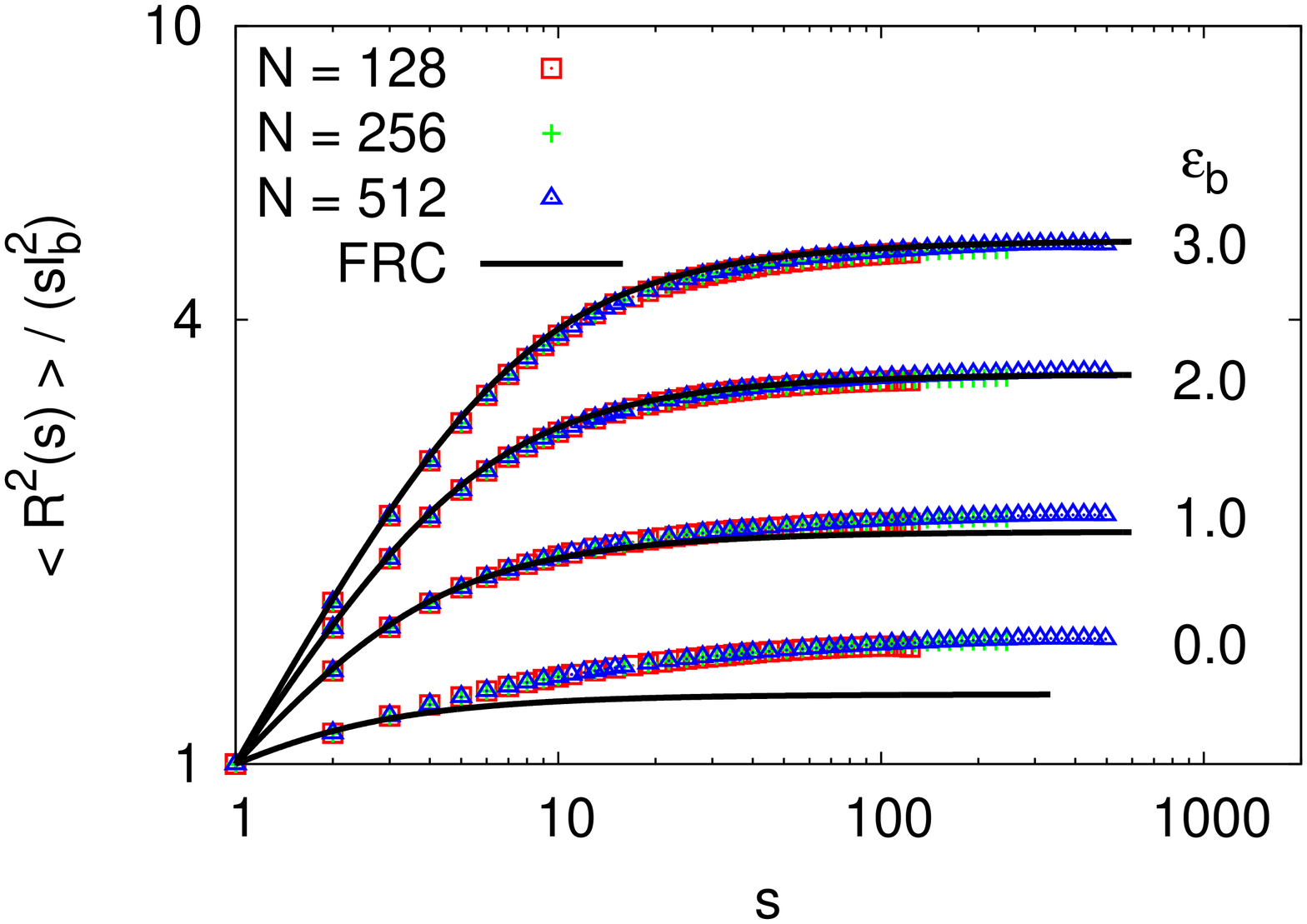}\\
\caption{(a) $\langle R^2(s) \rangle/s$ plotted versus $s$ for polymer melts
consisting of $n_cN=131072$ monomers, and for $\varepsilon_b=0.0$.
(b) $\langle R^2(s) \rangle/(s \ell_b^2)$ plotted versus $s$ for polymer melts
of $N=128$, $256$ and $512$ monomers, and for various values of bending energy $\varepsilon_b$,
as indicated. The theoretical predictions, Eq~(\ref{eq-rs2-meltf}) with
$b_e=3.244$ and $c_s=0.412$ in (a), and
Eq.~(\ref{eq-Rs2-FRC}) in (b), are shown by curves for comparison.}
\label{fig-melt-rs2}
\end{center}
\end{figure*}

\begin{figure*}[t]
\begin{center}
(a)\includegraphics[scale=0.29,angle=270]{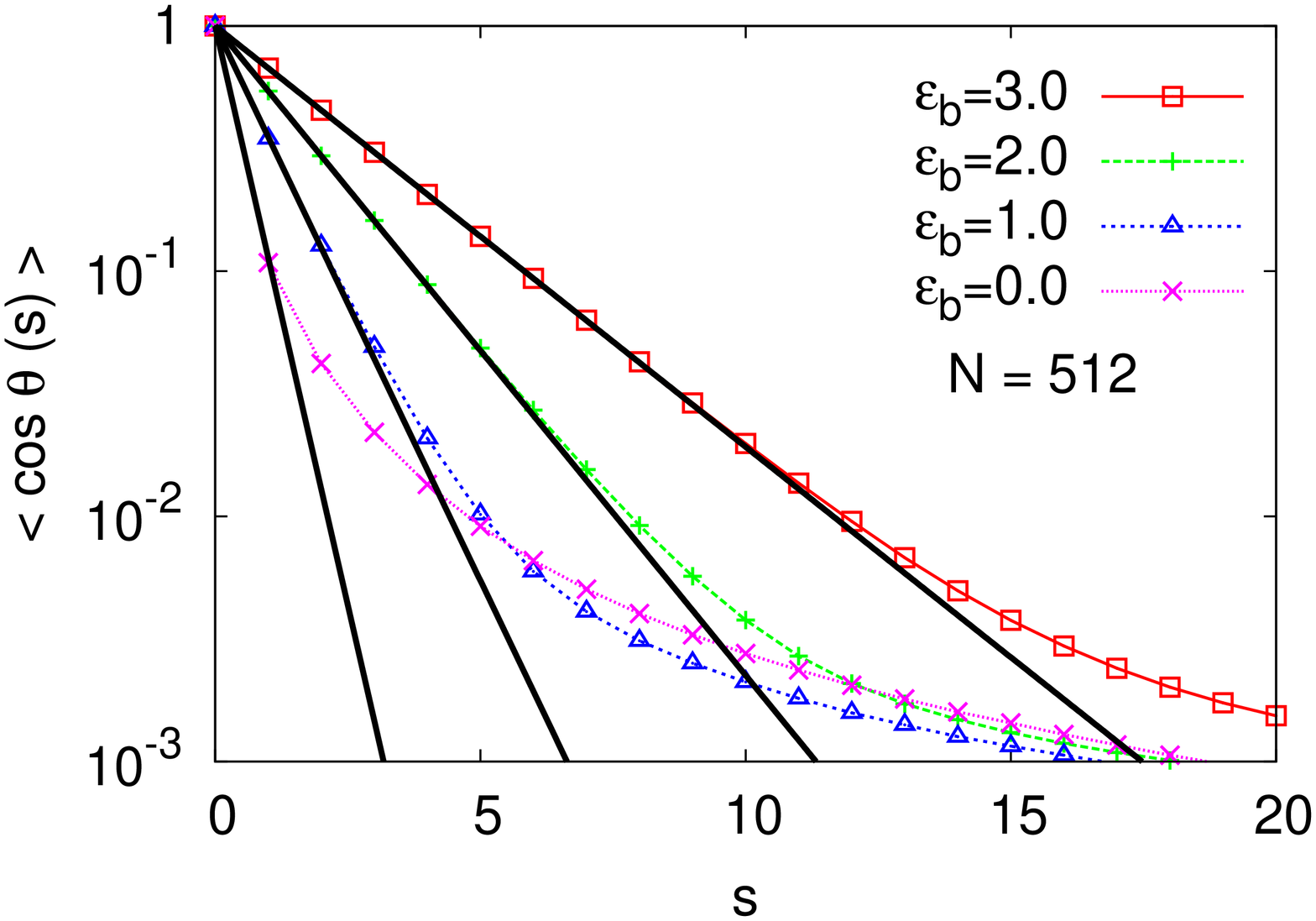} \hspace{0.4cm}
(b)\includegraphics[scale=0.29,angle=270]{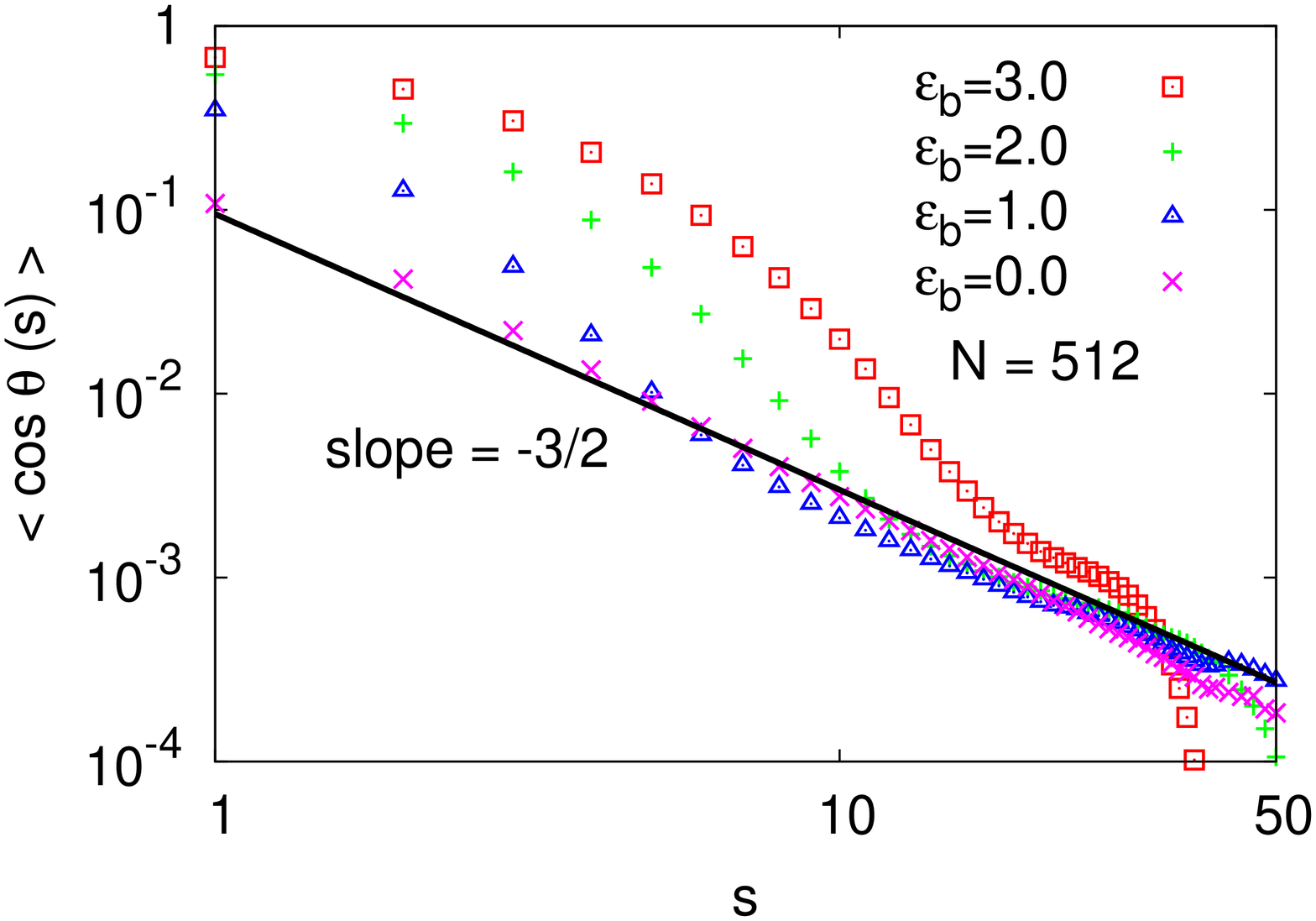}\\
\caption{(a) Semi-log plot of the bond-bond orientational correlation function
$\langle \cos \theta(s) \rangle$ vs.\ $s$.
{(b) Same data as shown in (a), but
on a log-log scale.}
Data are for polymer melts consisting of semiflexible chains of
size $N=512$, and for $\varepsilon_b=0.0$, $1.0$,
$2.0$ and $3.0$, as indicated. In (a) the straight lines indicate the fits of the initial decay,
$\langle \cos \theta(s) \rangle \propto \exp(-s\ell_b/\ell_p)$
(Eq.~(7)), for $\varepsilon_b>0$.
For $\varepsilon_b=0$ the straight line is described by $\exp(-s\ell_b/\ell_{p,\theta})$.
Values of $\ell_p/\ell_b$ and $\ell_{p,\theta}/\ell_b$ are listed in Table~III.
{In (b) the straight line indicates the power law,
$\langle \cos \theta(s) \rangle \propto s^{-3/2}$.}}
\label{fig-hs-melt}
\end{center}
\end{figure*}

\begin{table*}[htb]
\caption{Estimates of root-mean-square end-to-end distance $\langle R_e^2 \rangle^{1/2}$,
gyration radius $\langle R_g^2 \rangle^{1/2}$, and bond length
$\ell_b=\langle \vec{b}^2 \rangle^{1/2}$
for polymer melts containing $n_c$ fully flexible chains of $N$ monomers ($\varepsilon_b=0$).}
\label{table2}
\begin{center}
\begin{tabular}{|rrccc|}
\hline
$N$ & $n_c$ & $\langle R_e^2 \rangle^{1/2} $  & $\sqrt{6}\langle R_g^2 \rangle^{1/2}$ & $\ell_b$ \\
\hline
32  & 4096 &   17.07 &  17.13   & 2.63 \\
64  & 2048 &   24.74 &  24.75   & 2.63 \\
128 & 1024 &   35.54 &  35.51   & 2.63 \\
256 &  512 &   50.76 &  50.73   & 2.63 \\
512 &  256 &   72.17 &  72.20   & 2.63 \\
\hline
\end{tabular}
\end{center}
\end{table*}

In the equilibration process, each Monte Carlo step contains $n_cN$ L26 moves where
each monomer is selected once to move, $n_c$ slithering-snake moves where each
chain is selected once to move,  and $n_c$ pivot moves where each chain is also selected
once to move.
There are about $10^6$ independent configurations for each measurement in equilibrium.
Figure~\ref{fig-melt-rs2} represents the well-equilibrated data for polymer melts in
a box of size $L^3=128^3$. Results for fully flexible chains of 
chain sizes $N=32$, $64$, $128$,
and $512$ shown in Fig.~\ref{fig-melt-rs2}a are in perfect agreement with
the theoretical prediction given in Ref.~\cite{Wittmer2007,Wittmer2011},
\begin{equation}
  R^2(s) = b_e^2s(1-c_s/\sqrt{s}) \qquad {\rm for} \, 1 \ll s \ll N
\label{eq-rs2-meltf}
\end{equation}
where $b_e=\lim_{s \rightarrow \infty} (R^2(s)/s)^{1/2}$ is the 
``effective bond length" (which is different from the
root-mean-square bond length $\ell_b$ between monomers along a chain),
and $c_s\equiv \sqrt{24/\pi^3}/(\rho b_e^3)$.
Results for polymer melts of $N=128$, $256$, and $512$ monomers are shown
in Fig.~\ref{fig-melt-rs2}b for $\varepsilon_b=0.0$, $1.0$, $2.0$, and $3.0$.
The theoretical prediction for a freely rotating chain (FRC), 
Eq.~(\ref{eq-Rs2-FRC}), is also included for comparison. For $\varepsilon_b=0.0$ we see
a strong deviation from the theoretical prediction
since chains tend to swell due to the excluded volume effect
that causes the $1/\sqrt{s}$ correction in Eq.~(\ref{eq-rs2-meltf}).
This is different from the results~\cite{Auhl2003,Moreira2014} 
obtained using the bead-spring model in the continuum, where on short and intermediate length
scales Eq.~(\ref{eq-Rs2-FRC}) for a FRC overestimates the internal distances for fully
flexible chains, while on a large length scale it gives the accurate prediction.  
As the bending energy $\varepsilon_b$ increases,
we see in Fig.~\ref{fig-melt-rs2}b that the deviation from the prediction 
for a FRC reduces. 

We also estimate the root-mean-square bond length
$\ell_b=\langle \vec{b}^2 \rangle^{1/2}$, end-to-end distance
$\langle R_e^2 \rangle$, and radius of gyration $\langle R_g^2 \rangle$.
Results for various values of $\varepsilon_b$ and $N$
are listed in Table~\ref{table2} and \ref{table3}.
We see that $\langle R_e^2 \rangle / \langle R_g^2 \rangle \approx 6$ which 
is predicted for ideal chains. Our results of $\langle R_e^2 \rangle$,
$\langle R_g^2 \rangle$, and $\ell_b$ for $\varepsilon_b=0.0$ and for 
various values of $N$ are in perfect agreement
with the results given in Ref.~\cite{Wittmer2007} although the 
total number of monomers of polymer chains in a melt 
and the lattice size which we have chosen for our simulations both are eight times smaller.
The persistence lengths for polymer melts consisting of semiflexible chains
determined by fitting the exponential decay, Eq.~(\ref{eq-cos-dWLC1}), and
the correlation between two neighboring bonds, Eq.~(\ref{eq-lp-dWLC}), are
shown in Fig.~\ref{fig-hs-melt}a and also listed in Table~\ref{table3}
including 
Flory's characteristic ratio $C_\infty$ using Eq.~(\ref{eq-Re2-N}).
Note that the asymptotic decay of $\langle \cos \theta(s) \rangle$
with $s$ is not exponential as predicted by Eq.~(\ref{eq-cos-dWLC1}), but
rather a power law decay, 
{$\langle \cos \theta(s) \rangle \propto s^{-3/2}$ for
$s^*< s \ll N$}, is expected, 
due to excluded volume effects~\cite{Wittmer2007,Hsu2010}.

Therefore only the initial decay of $\langle \cos \theta(s) \rangle$ with
$s$ {($s \ll s^*$)} is meaningful for the estimation of 
the persistence length.
{
In Fig.~\ref{fig-hs-melt}b we plot $\langle \cos \theta(s) \rangle$
versus $s$ on a log-log scale. The power law $s^{-3/2}$ is also shown
for comparison.  
We see that the range over which the power law 
behavior holds ($s > s^*$) shrinks with increasing chain stiffness,
and the data deviate from the straight line showing that
the finite-size effect sets in.}
Results for various values of $\varepsilon_b$ and for chain sizes
$N=128$ and $N=512$ are listed in Table~\ref{table3}.
The relationship $\ell_p/\ell_b \approx \ell_{p,\theta}/\ell_b \approx C_\infty/2$
holds for semiflexible chains ($\varepsilon_b>0$) as predicted in 
Eqs.~(\ref{eq-Re2-N}) and (\ref{eq-Re2-FRC}). 
The root-mean-square bond length $\ell_b \approx 2.63$ for polymer melts, shown
in Table~\ref{table2}, is smaller than $\ell_b \approx 2.72$ for linear flexible chains
in dilute solution based on the BFM.
The reason is that the crowded environment of polymer chains
prevents the bond vector between two effective monomers going through several
unit cells, i.e. occupying large volume.
We also see that $\ell_b$ depends only weakly on the stiffness of the chains
as shown in Table~\ref{table3}.

\begin{table*}[htb]
\caption{Estimates of root-mean-square end-to-end distance $\langle R_e^2 \rangle^{1/2}$,
gyration radius $\langle R_g^2 \rangle^{1/2}$, bond length 
$\ell_b=\langle \vec{b}^2 \rangle^{1/2}$,
persistence lengths, $\ell_p/\ell_b$ and $\ell_{p,\theta}/\ell_b$,
using Eqs.~(\ref{eq-cos-dWLC1}) and ~(\ref{eq-lp-dWLC}), respectively, 
Flory's characteristic ratio from Eq.~(\ref{eq-Re2-N}),
and the dimensionless compressibility $C_g=\rho k_BT\kappa_T$ from Eq.~(\ref{eq-kappa})
for semiflexible chains in a melt for various values of bending energy 
$\varepsilon_b$, and for $N=512$ and $N=128$.}
\label{table3}
\begin{center}
\begin{tabular}{|c|cccc|cccc|}
\hline
$N$ & \multicolumn{4}{|c|}{512} & \multicolumn{4}{|c|}{128} \\
\hline
$\varepsilon_b$ & 0.0 & 1.0 & 2.0 & 3.0 & 0.0 & 1.0 & 2.0 & 3.0\\
\hline
$\langle R_e^2 \rangle^{1/2}$ & 72.17 & 87.25 & 108.95 & 132.82 & 35.54 & 42.96 & 53.59 & 65.38\\
$\sqrt{6}\langle R_g^2 \rangle^{1/2}$ & 
72.20 & 87.12 & 108.69 & 132.48 & 35.51 & 42.84 & 53.20 & 64.41\\
$\ell_b$ & 2.63 & 2.62 & 2.62 & 2.61 & 2.63 & 2.62 & 2.62 & 2.61\\
$\ell_p/\ell_b$ & \_ & 0.96 & 1.64 & 2.53  & \_ & 0.96 & 1.64 & 2.53\\
$\ell_{p,\theta}/\ell_b$ & 
0.45 & 0.95 & 1.64 & 2.53 & 0.45 & 0.95 & 1.64 & 2.53 \\
$C_\infty$ & 1.24 & 2.06 & 3.38 & 5.12 & 1.24 & 2.06 & 3.38 & 5.12 \\
$C_g$ & 0.25 & 0.24 & 0.23 & 0.23  & 0.25 & 0.24 & 0.23 & 0.23\\
\hline
\end{tabular}
\end{center}
\end{table*}

\begin{figure*}[t]
\begin{center}
(a)\includegraphics[scale=0.29,angle=270]{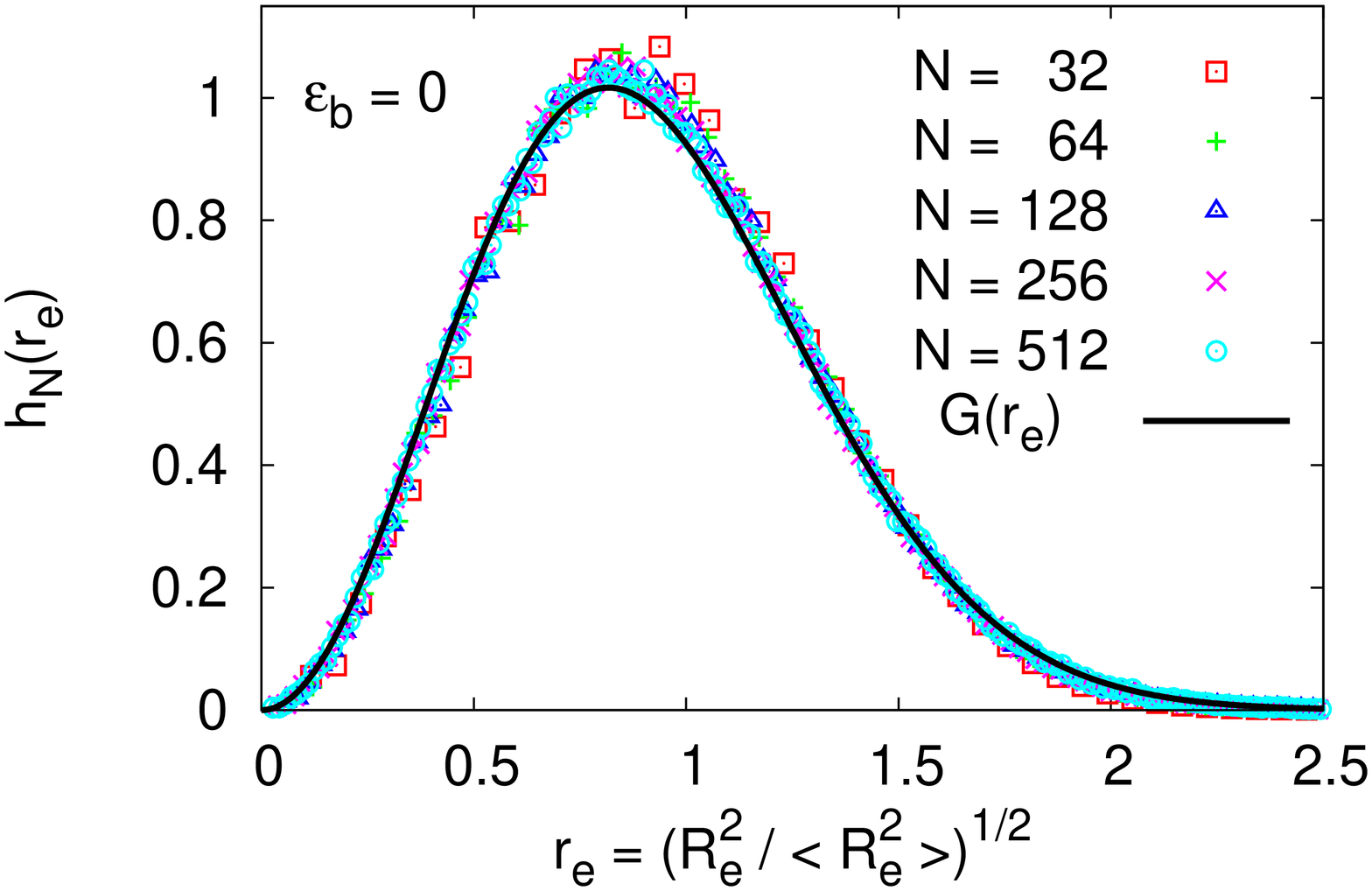} \hspace{0.4cm}
(b)\includegraphics[scale=0.29,angle=270]{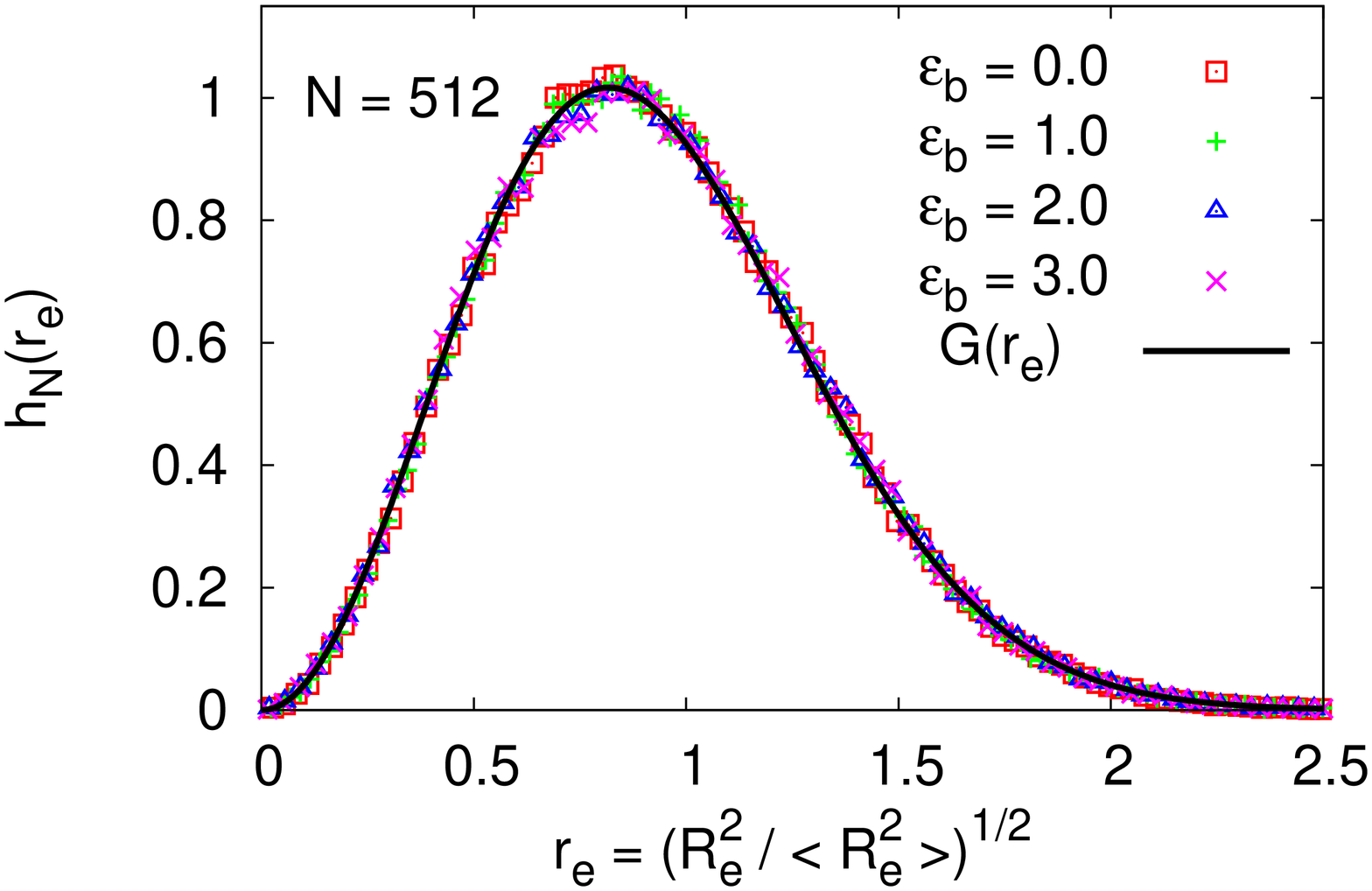}\\
\caption{Normalized probability distributions of
$r_e=(R_e^2/\langle R_e^2 \rangle)^{1/2}$, $h_N(r_e)$, plotted
versus $r_e$ for polymer chains in a melt.
In (a) data are for $\varepsilon=0.0$, and for various values of $N$, as indicated.
In (b) data are for $N=512$, and for various values of $\varepsilon_b$, as indicated.
The theoretical prediction $G(r_e)$, \{Eq.~(\ref{eq-pre})\}, (solid curve) is also
shown for comparison.}
\label{fig-pRe}
\end{center}
\end{figure*}

\begin{figure*}[ht]
\begin{center}
(a)\includegraphics[scale=0.29,angle=270]{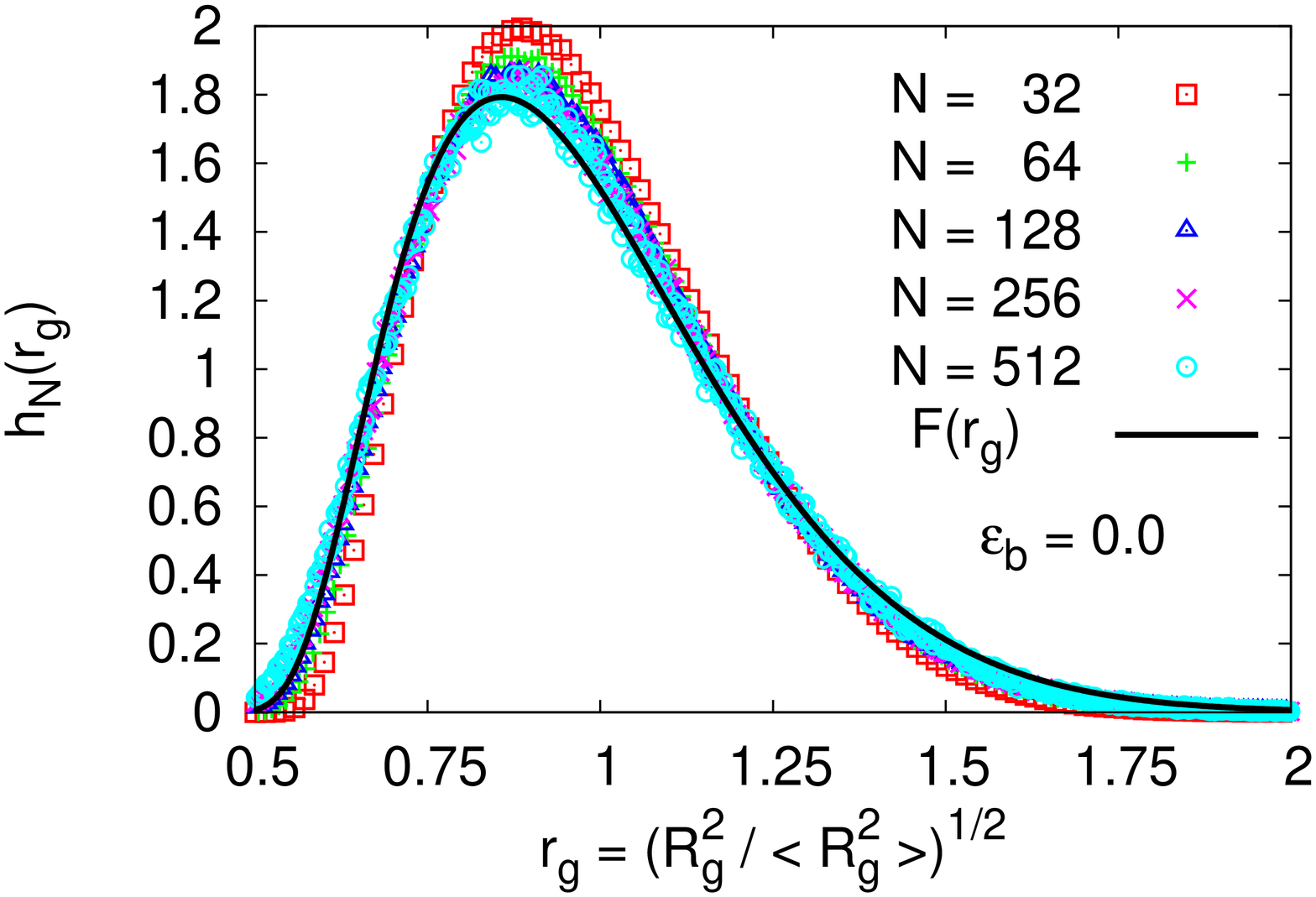} \hspace{0.4cm}
(b)\includegraphics[scale=0.29,angle=270]{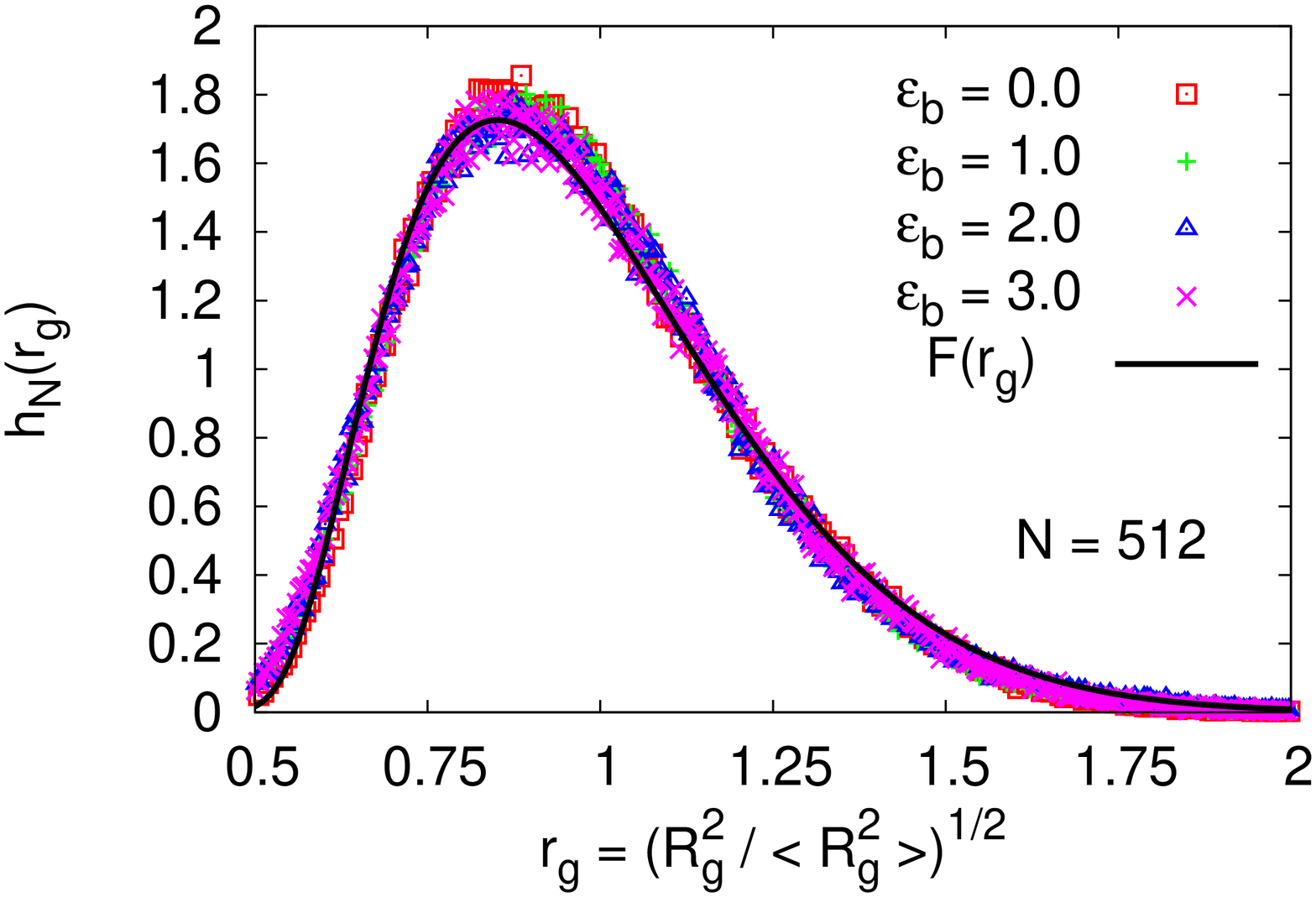}\\
(c)\includegraphics[scale=0.29,angle=270]{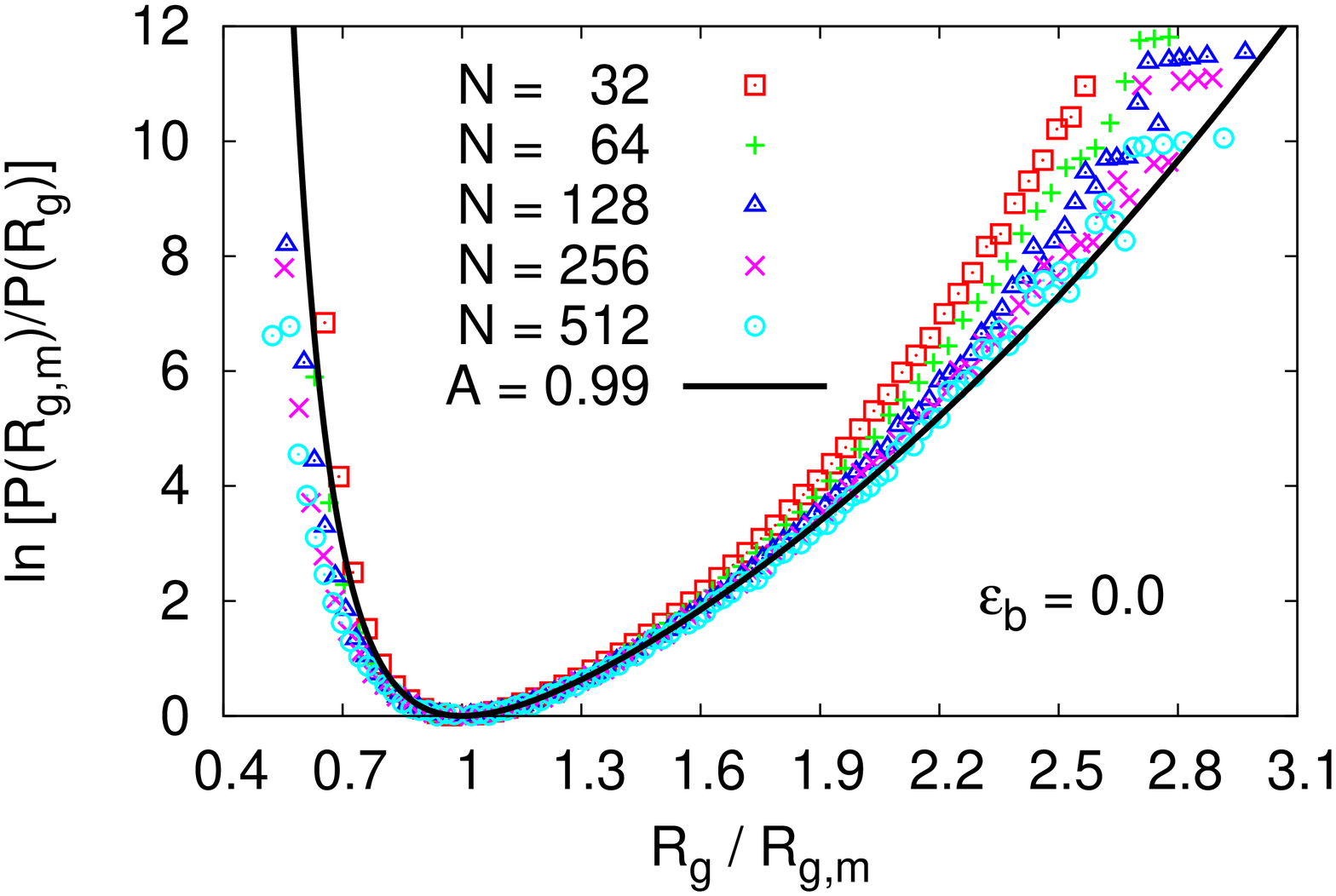} \hspace{0.4cm}
(d)\includegraphics[scale=0.29,angle=270]{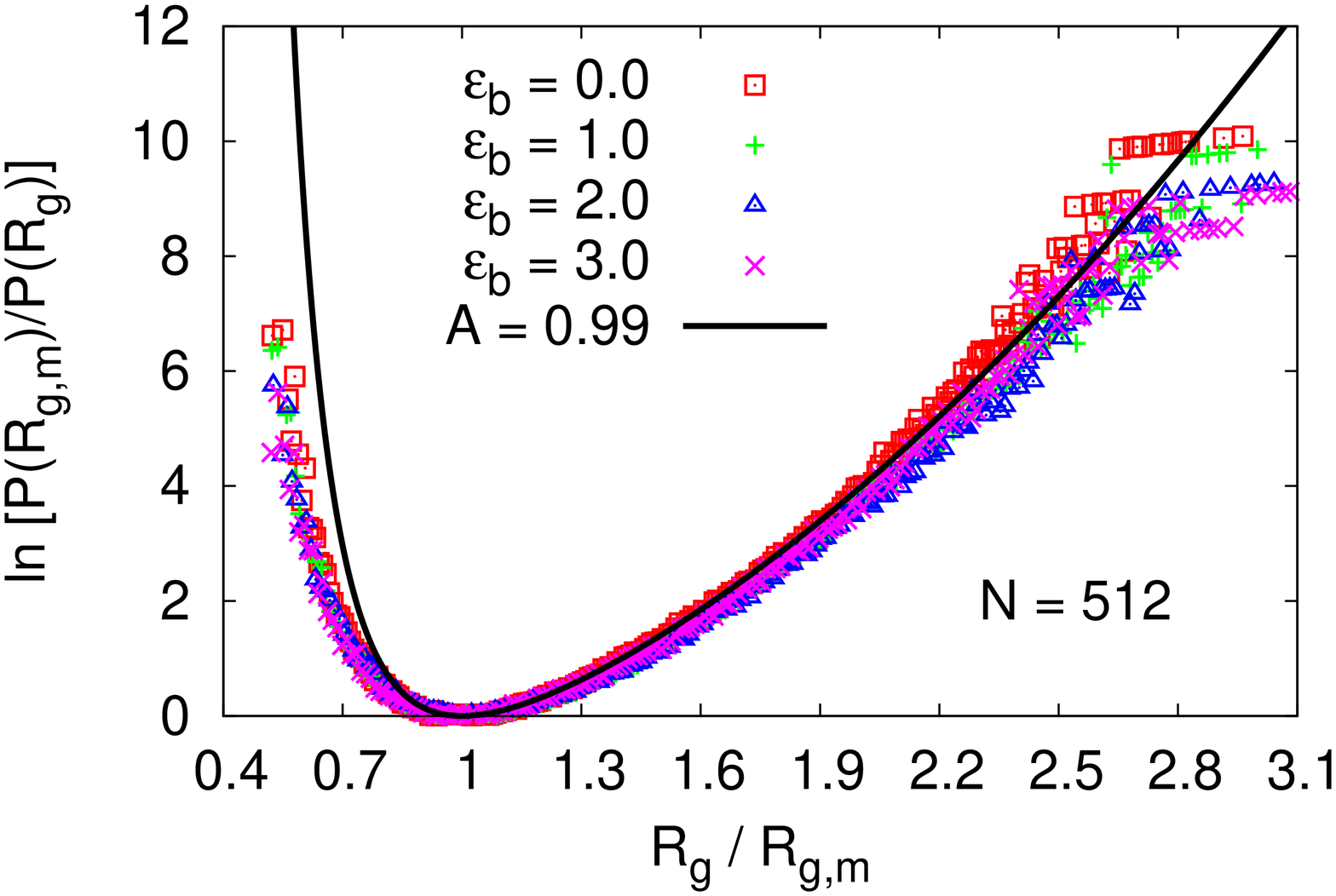}\\
\caption{(a)(b) Normalized probability distributions of
$r_g=(R_g^2/\langle R_g^2 \rangle)^{1/2}$, $h_N(r_g)$, plotted
versus $r_g$ for polymer chains in a melt.
(c)(d) Logarithm of the rescaled probability distribution of gyration
radius, $\ln(P(R_{g,m})/P(R_g))$ as a function of $R_g/R_{g,m}$.
In (a)(c) data are for $\varepsilon_b=0.0$, and for various values of $N$, as indicated.
The theoretical prediction $F(r_g)$, Eq.~(\ref{eq-prg}), (solid curve) is also shown
for comparison.
In (b)(d) data are for $N=512$, and for various values of $\varepsilon_b$, as indicated.
The theoretical prediction $F(r_g)$, Eq.~(\ref{eq-A}), with $A=0.99$ (solid curve) is also
shown for comparison.}
\label{fig-pRg}
\end{center}
\end{figure*}

\subsection{Probability distributions of $R_e$ and $R_g$}

The conformational behavior of polymer chains of size $N$ 
in a melt can also be described by 
the probability distributions of end-to-end distance $\vec{R}_e$ 
and gyration radius $R_g$, 
$P_N(\vec{R}_e)$ and $P_N(R_g)$, respectively. 
The probability distribution of $\vec{R}_e$ for ideal chains is simply
a Gaussian distribution,
\begin{equation}
    P_N(\vec{R}_e)=\left(\frac{3}{2\pi N\ell_b^2}\right)^{3/2}
\exp \left(-\frac{3R_e^2}{2N\ell_b^2} \right) \,.
\label{eq-pRe}
\end{equation} 
There exists exact theoretical prediction of the 
probability distribution 
of $R_g$ for ideal chains~\cite{Yamakawa1970,Fujita1970,Denton2002},
but they are complicated to evaluate. 
However, it has been checked~\cite{Vettorel2010,Hsu2014,Froehlich2013}
that the same formula suggested by Lhuillier~\cite{Lhuillier1988} for 
polymer chains under good solvent conditions in $d$-dimensions is still
a good approximation for ideal chains, i.e.,
\begin{equation}
 P_N(R_g) \sim \exp \left[-a_1 \left(\frac{N^\nu}{R_g}\right)^{\alpha d}
-a_2 \left(\frac{R_g}{N^\nu} \right)^\delta \right]
\label{eq-pRg}
\end{equation}
where $a_1$ and $a_2$ are (non-universal) constants, and the exponents $\alpha$ and
$\delta$ are linked to the space dimension $d$ and the Flory exponent $\nu$ by
$\alpha=(\nu d-1)^{-1}$ and $\delta=(1-\nu)^{-1}$. Here $(1+\alpha)$
is the des Cloizeaux exponent~\cite{Cloizeaux1975} for the osmotic pressure of
a semidilute polymer solution, and $\delta$ is the Fisher exponent~\cite{Fisher1966}
characterizing the end-to-end distance distribution.

Numerically,
the probability distribution of $x$ is obtained by accumulating the 
histogram $H_N(x)$ of $x$ over all 
configurations and all chains of size $N$, given by
\begin{equation}
     H_N(x) = \sum_{\rm config.} \delta_{x,x'}
\end{equation}
here $x$ stands for $R_e$ or $R_g$.
Note that an angular average over all directions has been included in the accumulating
process of the histogram due to spherical symmetry. Thus, the normalized histogram
of $x$ is given by
\begin{equation}
    h_N(x)=H_N(x)/\sum_{x'} H_N(x')=4 \pi C_N x^2 P_N(x)
\end{equation}
where $C_N$ is the normalization factor such that
\begin{equation}
      C_N \int_0^\infty 4 \pi x^2 P_N(x) =1 \,.
\end{equation}

In order to compare the probability distributions of $R_e$ 
and $R_g$ between various different chain sizes $N$ and bending energies 
$\varepsilon_b$, instead of $R_e$ and $R_g$ we use the reduced
end-to-end distance $r_e=(R_e^2 /\langle R_e^2 \rangle)^{1/2}$ and
the reduced gyration radius $r_g=(R_g^2 /\langle R_g \rangle)^{1/2}$. 
Results of the normalized histogram $h_N(r_e)$ compared to the theoretical prediction 
\begin{equation}
     G(r_e)= 4 \pi r_e^2 C_N \left(\frac{3}{2\pi}\right)^{3/2}
\exp \left(-\frac{3r_e^2}{2} \right)
\label{eq-pre}
\end{equation}
using $\langle R_e^2 \rangle \sim N$ and Eq.~(\ref{eq-pRe}) are
shown in Fig.~\ref{fig-pRe}. Here the normalization factor $C_N=1$ for 
all cases.
We see the nice data collapse for chain sizes $N \ge 64$ and all
bending energies $\varepsilon_b$ in Fig.~\ref{fig-pRe}a,b and the data
are described by a universal scaling function $G(r_e)$ for ideal chains.
For $N=32$ the data deviates from the master curve due to the finite-size
effect. Results of the normalized histogram $h_N(r_g)$ compared to the 
theoretical prediction
\begin{equation}
     F(r_g)= 4 \pi r_g^2 C_N \exp \left(-b_1 r_g^{-\alpha d} -b_2 r_g^{\delta}\right)
\label{eq-prg}
\end{equation}
using $\langle R_g^2 \rangle = k_g^2 N$ where $k_g$ is a constant and Eq.~(\ref{eq-pRg}) are
shown in Fig.~\ref{fig-pRg}. Here the normalization factor $C_N$, parameters
$b_1= a_1 k_g^{-\alpha d}$ and $b_2 = a_2 k_g^\delta$ are determined numerically depending 
on the simulation data, and they 
depend slightly on the chain size $N$. As $N \rightarrow \infty$ they will 
tend to asymptotic values (not shown). 
Near the peak ($r_g \approx 0.85$), we see the finite-size effect is 
stronger for the distribution of $r_g$.
The fitting curve $F(r_g)$ with parameters
$b_1=0.09$, $b_2=2.29$, and $C_N = 1.31$ determined by the least-square fit
for $N=512$ and $\varepsilon_b=0.0$ and 
$b_1=0.075$, $b_2=2.20$, and $C_N=1.13$ for $N=512$ and $\varepsilon_b=3.0$
are shown in Fig.~\ref{fig-pRg}a,b, respectively for comparison. 
The distribution $P(R_g) \propto H_N/R_g^2$ has its maximum value,
i.e. $P(R_g=R_{g,m})=\max\; P(R_g)$, and the corresponding gyration radius
$R_{g,m} \propto R_g \propto N^{\nu}$. Therefore, using Eq.~(\ref{eq-pRg}),
the logarithm of the rescaled probability distribution as a function
of $(R_g/R_{g,m})$ is given by
\begin{eqnarray}
&&f \left( \frac{R_{g,m}}{R_g} \right) = \ln 
\frac{P(R_{g,m})}{P(R_g)} \nonumber \\
&=& A\left[\frac{1}{\alpha} \left( \frac{R_{g,m}}{R_g} \right)^{\alpha d}
+\frac{d}{\delta} \left( \frac{R_g}{R_{g,m}}\right)^\delta+1-d \right]
\label{eq-A}
\end{eqnarray}
with
\begin{equation}
  a_1=\frac{A}{\alpha} \left(\frac{R_{g,m}}{\ell_bN^\nu} \right)^{\alpha d}
\quad {\rm and} \quad a_2=\frac{Ad}{\delta} 
\left(\frac{\ell_b N^\nu}{R_{g,m}} \right)^\delta  \,.
\end{equation}
where one fitting parameter $A$ is left.
Results of $\ln \{P(R_{g,m})/P(R_g)\}$ plotted versus 
$R_g/R_{g,m}$ are presented in Fig.~\ref{fig-pRg}c,d.
We see that due to the finite-size effect, data for fully flexible chains in a melt 
shown in Fig.~\ref{fig-pRg}c,d only 
start to converge for $N>128$. As $N$ increases, the systematical errors need to
be taken into account. Using the least square fit, it gives $A=0.99(3)$.
It is obvious that the distribution can only
be well described by Eq.~(\ref{eq-A}) for $R_g>R_{g,m}$, similar as that 
was found for Gaussian chains in Ref.~\cite{Hsu2014}. 
For polymer melts of a fixed chain length $N=512$, we see that for $R_g<R_{g,m}$
the distribution remains the same for different stiffnesses, while for $R_g>R_{g,m}$
the data for $\varepsilon_b=0$ are slightly deviated from the data for $\varepsilon_b>0$.  
We plot the same distribution, Eq.~(\ref{eq-A}) with $A=0.99$ for comparison.

\subsection{Structure factor and compressibility}
\label{structure}
The collective static structure factor $S(q)$ of polymer melts is defined by
the total scattering from the center of all monomers inside the box, 
regardless of whether they are linked along a polymer chain or not, 
\begin{equation}
S(q)  
=\frac{1}{N_{\rm tot}} \left \langle \sum_{i=1}^{N_{\rm tot}} 
\sum_{j=1}^{N_{\rm tot}} \exp[i\vec{q} \cdot (\vec{r}_i-\vec{r}_j)] \right \rangle \,.
\end{equation}
Here $N_{\rm tot}=Nn_c$ are the total number of monomers and
$\langle \cdots \rangle$ represents the average over all independent
configurations, and over all vectors $\vec{q}$ of the same size
$q=\mid \vec{q} \mid$. Note that since a simple cubic lattice of size $L^3$ with
periodic boundary condition is considered for our simulations, only the following
$\vec{q}$ are allowed,
\begin{equation}
    \vec{q}=\frac{2\pi}{L}(n_1,n_2,n_3) 
\end{equation}
where $n_i=0,\pm 1,\pm 2,\ldots$ for $i=1$, $2$, and $3$, and $q\ge 2 \pi/L$.
$S(q)$ characterizes a competition between the intramolecular fluctuations 
($S_{\rm intra}$) and
the intermolecular correlations ($S_{\rm inter}$), thus one can write
\begin{equation}
 S(q)=S_{\rm intra}(q)+S_{\rm inter}(q)
\end{equation}
with
\begin{equation}
  S_{\rm intra}=\frac{1}{n_c}\left\langle \sum_{n=1}^{n_c}\frac{1}{N}\sum_{i,j=1}^N
\exp[i\vec{q} \cdot (\vec{r}_{i}^n-\vec{r}_{j}^n) ] \right\rangle =S_c(q)
\end{equation}
and
\begin{equation}
  S_{\rm inter} = \frac{1}{N_{\rm tot}}\sum_{i,j=1}^N \left \langle 
\sum_{n=1, n\neq n'}^{n_c} \sum_{n'=1}^{n_c}
\exp[i\vec{q} \cdot (\vec{r}_{i}^n -\vec{r}_{j}^{n'}) ] \right \rangle 
\end{equation}
where the contributions from the intramolecular fluctuations are
simply equivalent to the average of the standard static structure
factor of single polymer chains in a melt, $S_c(q)$.

\begin{figure*}[t]
\begin{center}
(a)\includegraphics[scale=0.29,angle=270]{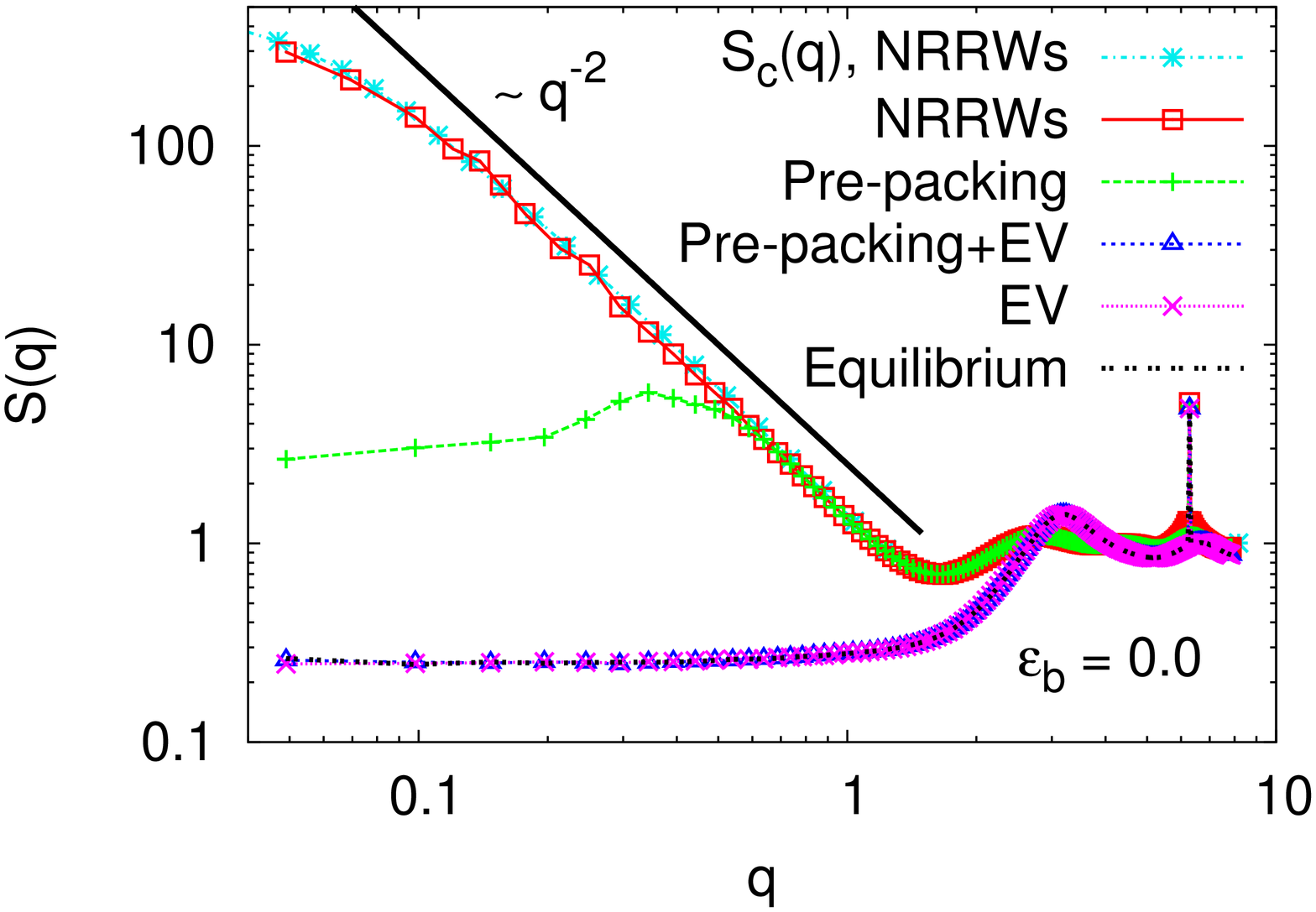} \hspace{0.4cm}
(b)\includegraphics[scale=0.29,angle=270]{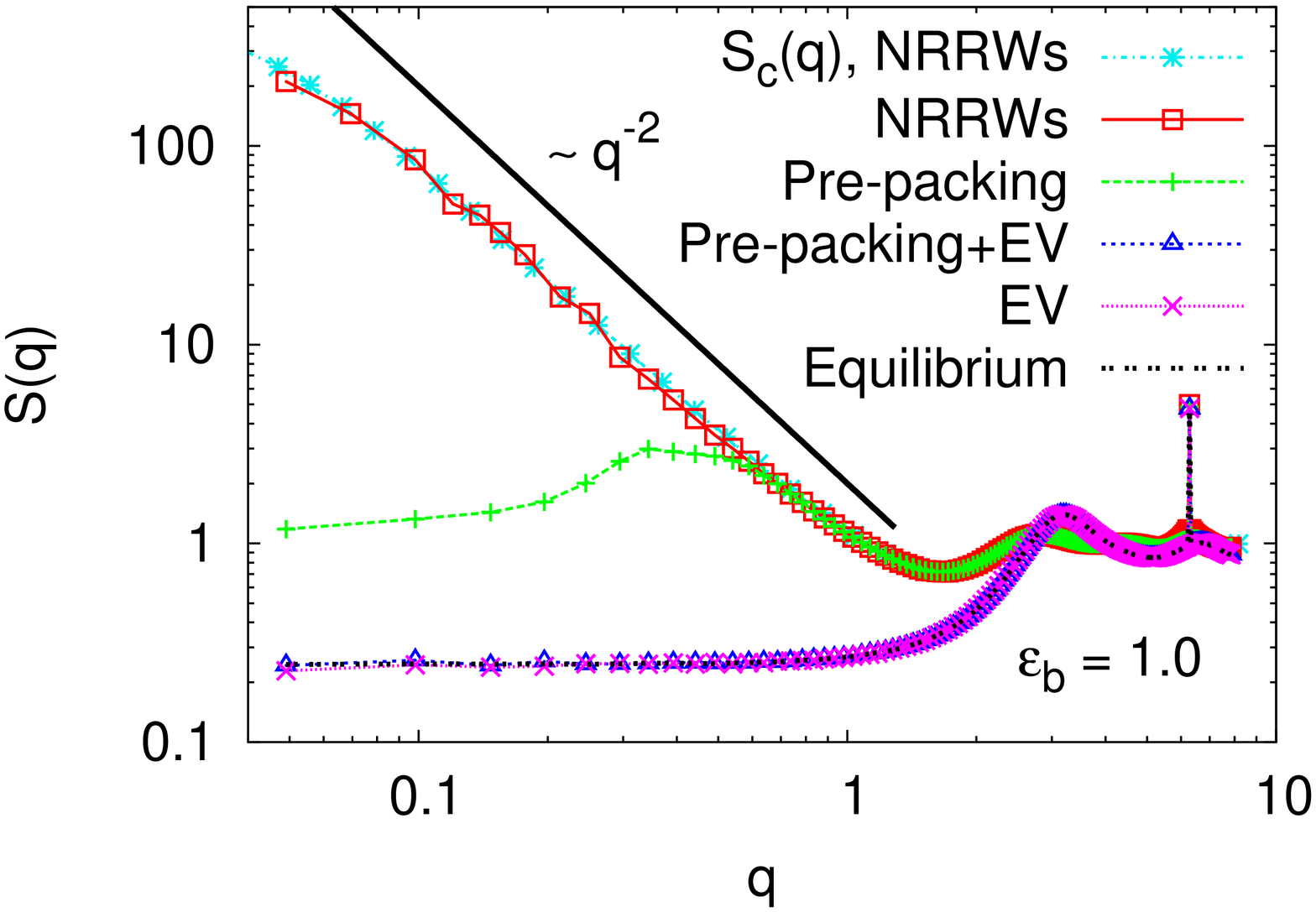}\\
(c)\includegraphics[scale=0.29,angle=270]{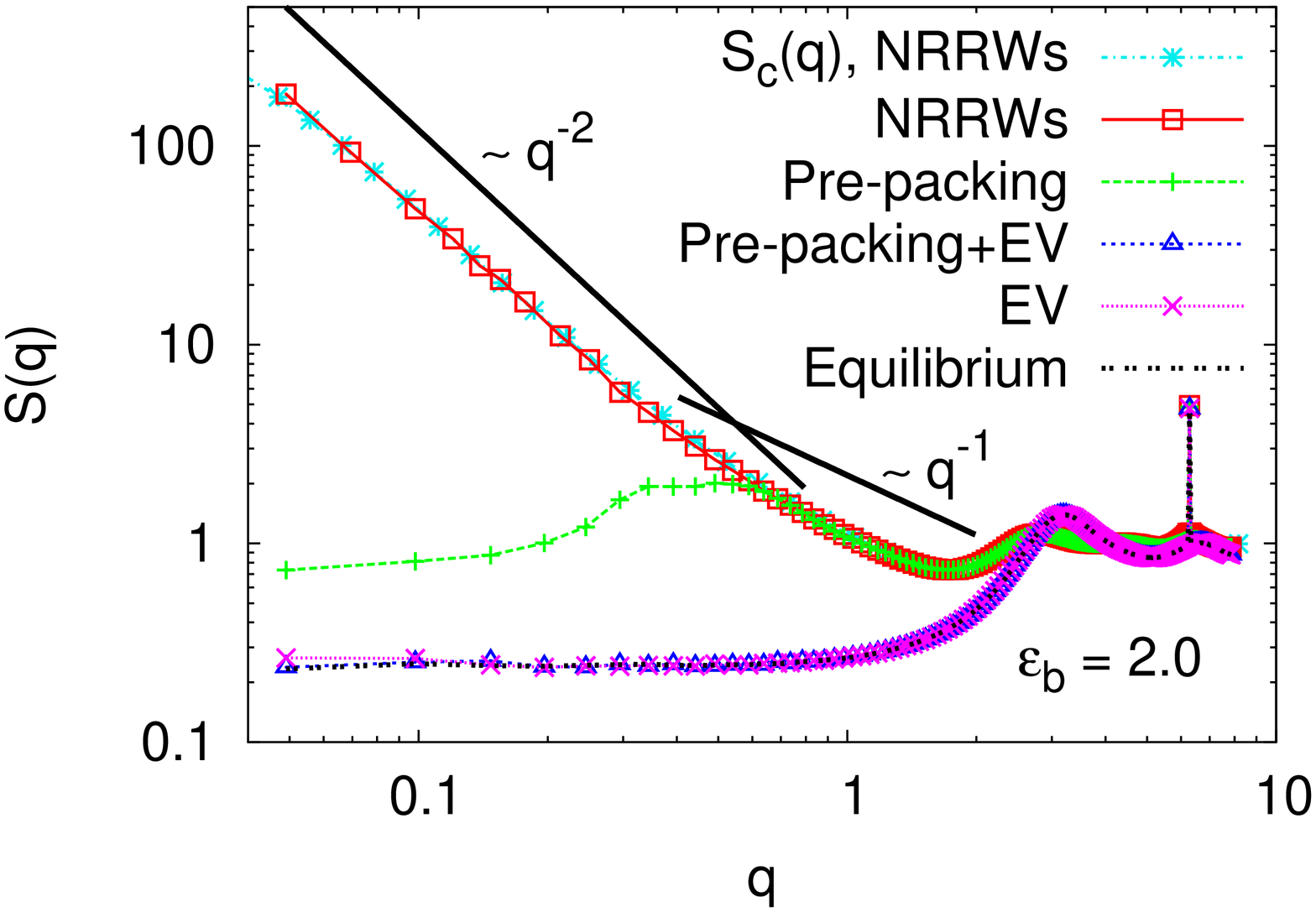} \hspace{0.4cm}
(d)\includegraphics[scale=0.29,angle=270]{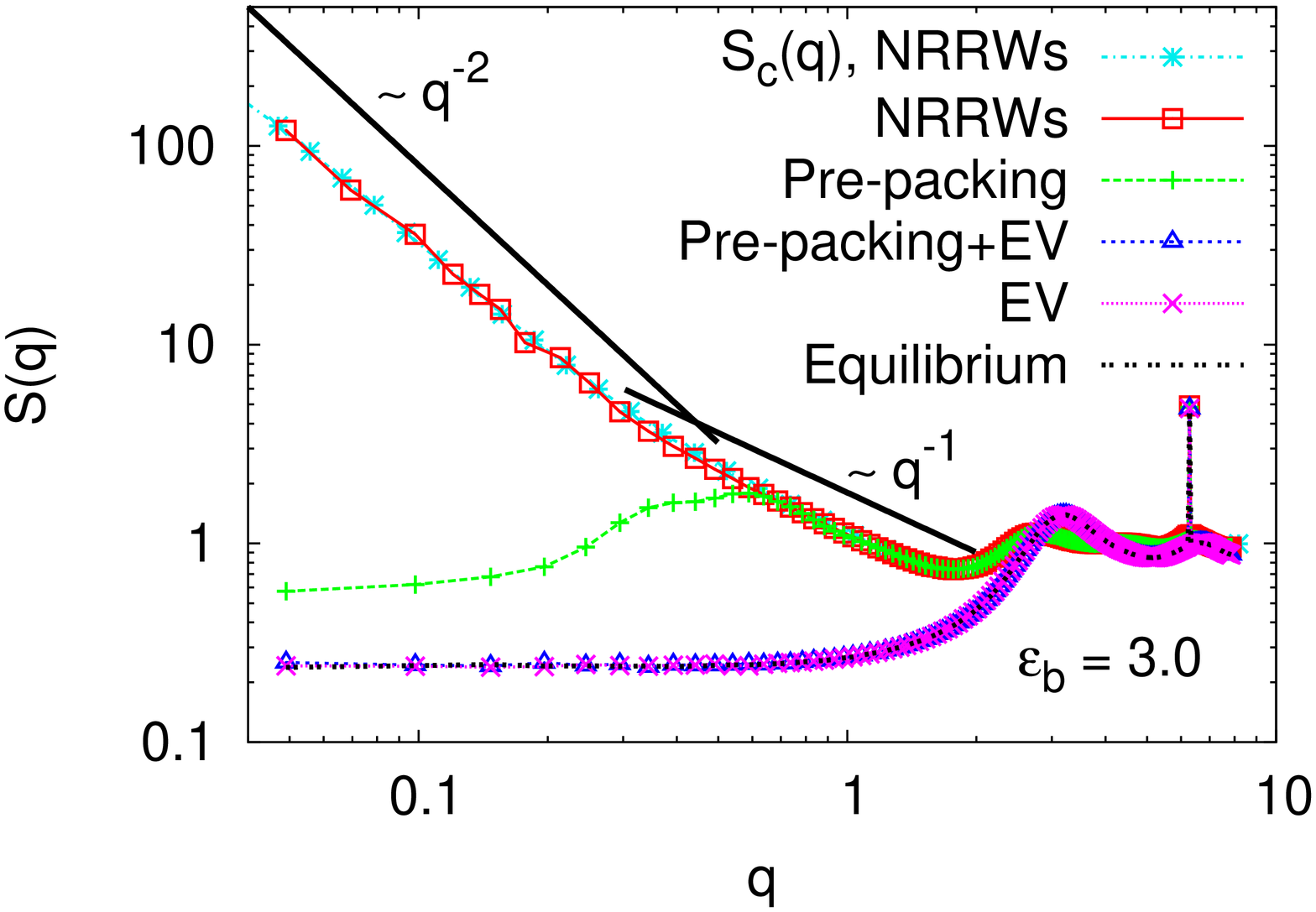}\\
\caption{Structure factor $S(q)$ from the scattering of polymer
melts plotted vs.\ $q$ on log-log scales for $\varepsilon_b=0.0$ (a), $1.0$ (b),
$2.0$ (c) and $3.0$ (d).
Results are for the conformations of polymer melts at different stages, initial NRRWs (NRRWs), 
rearranged NRRWs after applying pre-packing process (Pre-packing), 
SAWs obtained from NRRWs by the push-off process before applying the pre-packing procedure
(EV), and after (Pre-packing+EV), and equilibrated SAWs after polymer melts reaching
equilibrium (Equilibrium). The structure factor $S_c(q)$ for single NRRWs chains
generated initially and the Scaling predictions for Gaussian coils, 
$S(q)\sim q^{-2}$, and for rigid rods, $S(q) \sim q^{-1}$, are also shown
for comparison.}
\label{fig-sq-melt512i}
\end{center}
\end{figure*}

The collective structure factors $S(q)$ which represent the conformations of the
whole polymer system at different stages from weakly-interactive NRRW chains to
equilibrated SAW chains in a melt for $N=512$ and for $\varepsilon_b=0.0$,
$1.0$, $2.0$, and $3.0$ are shown in Fig.~\ref{fig-sq-melt512i}.
Results include the initial NRRW chains (NRRWs), rearranged NRRW chains through
the pre-packing process (Pre-packing), initial SAW chains obtained from the initial
NRRW chains by switching on the excluded volume effect before (EV) and
after (Pre-packing+EV) applying the pre-packing procedure, and
the equilibrated SAW chains in a melt (Equilibrium).
Each curve shows the result averaged over $32$ independent configurations at
the intermediate states, while $1500$ independent configurations
are considered for the measurement in equilibrium.
The average static structure factor of single NRRW chains generated initially,
$S_c(q)$, is also included for comparison.
In Fig.~\ref{fig-sq-melt512i} we see that in all cases there is no interactions
between NRRWs generated in the box at the beginning,
so the structure factor shows the same scaling behavior for single NRRWs
in dilute solution, i.e. $S(q)=S_c(q)$.
The scaling predictions $S(q) \sim q^{-1/\nu_{id}}$ with $\nu_{id}=1/2$ for ideal
chains and $S(q) \sim q^{-1}$ for rigid rods are verified for fully flexible and
moderately stiff chains.
After the pre-packing process where NRRW chains are rearranged, the collective
structure factor decreases in the low-q regime.
Once all overlapping monomers are pushed off there is no difference between
the configurations generated by the push-off process with and without the pre-packing
in the intermediate and large $q$, but for very small $q$ a small discrepancy
exists between those data sets. It is consistent with our previous observation
of $\langle R^2(s) \rangle$ in Fig.~\ref{fig-rs-melt512} that the discrepancy
becomes more prominent at larger scales.

\begin{figure}[t]
\begin{center}
\includegraphics[scale=0.29,angle=270]{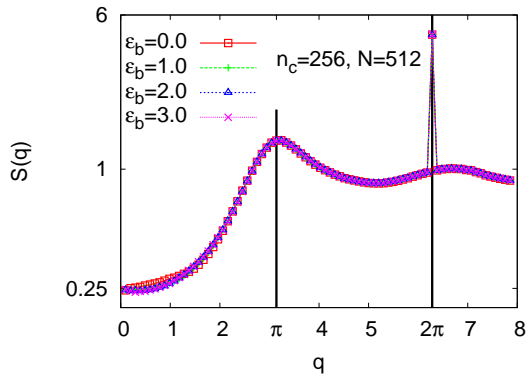} \hspace{0.4cm}
\caption{Structure factors $S(q)$ of polymer melts, plotted vs.\ $q$
on semi-log scales
Data are for polymer melts consisting of $n_c=256$ chains of lengths $N=512$, 
and for various values of $\varepsilon_b$, as indicated. 
In (a) the peak appears at $q\approx \pi$  measuring the mean inter-particle 
distance in the polymer melts, and the second peak at $q\approx 2\pi$ 
is the so-called Bragg peak.} 
\label{fig-sq-melt512}
\end{center}
\end{figure}

\begin{figure*}[t]
\begin{center}
(a)\includegraphics[scale=0.29,angle=270]{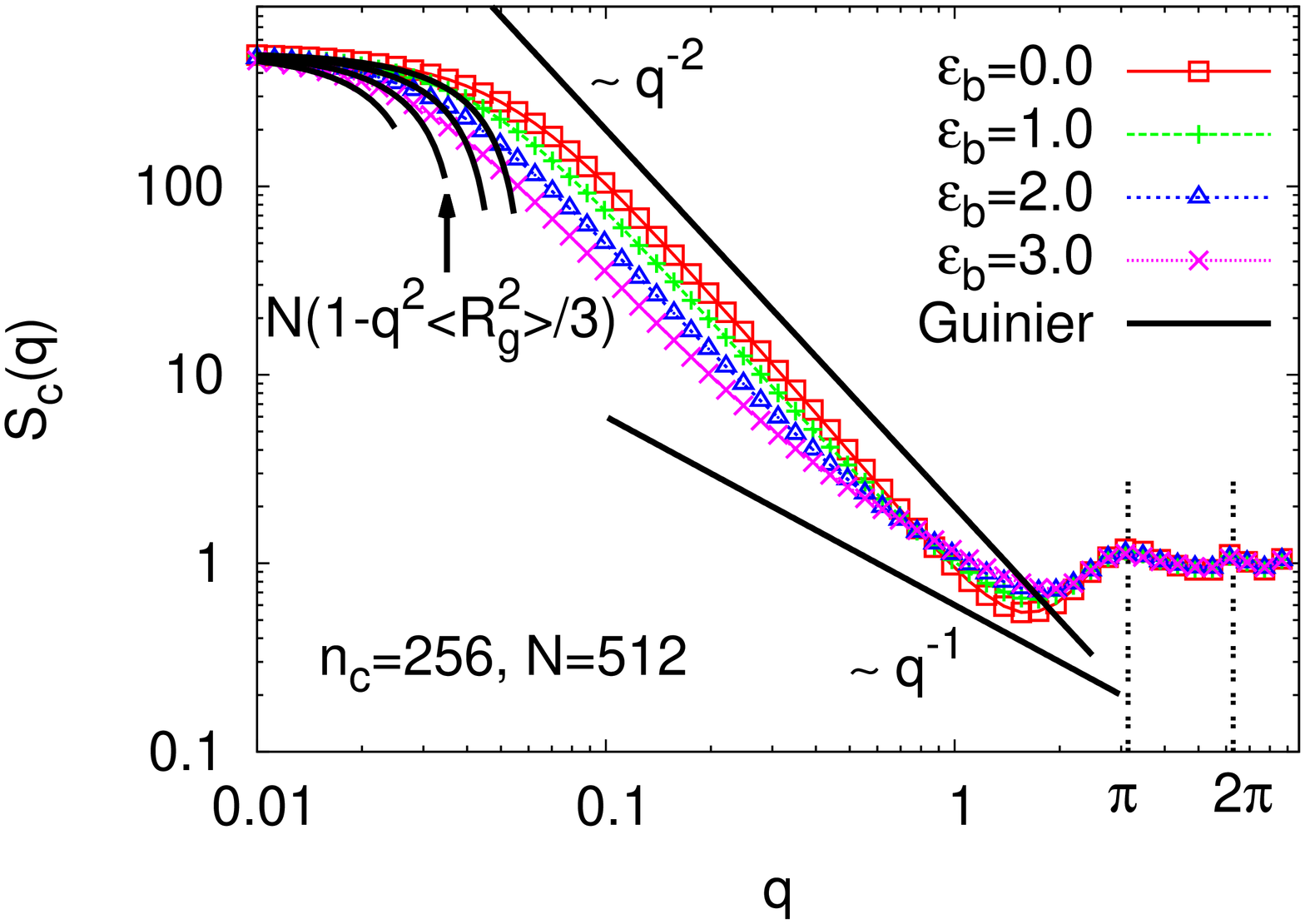} \hspace{0.4cm}
(b)\includegraphics[scale=0.29,angle=270]{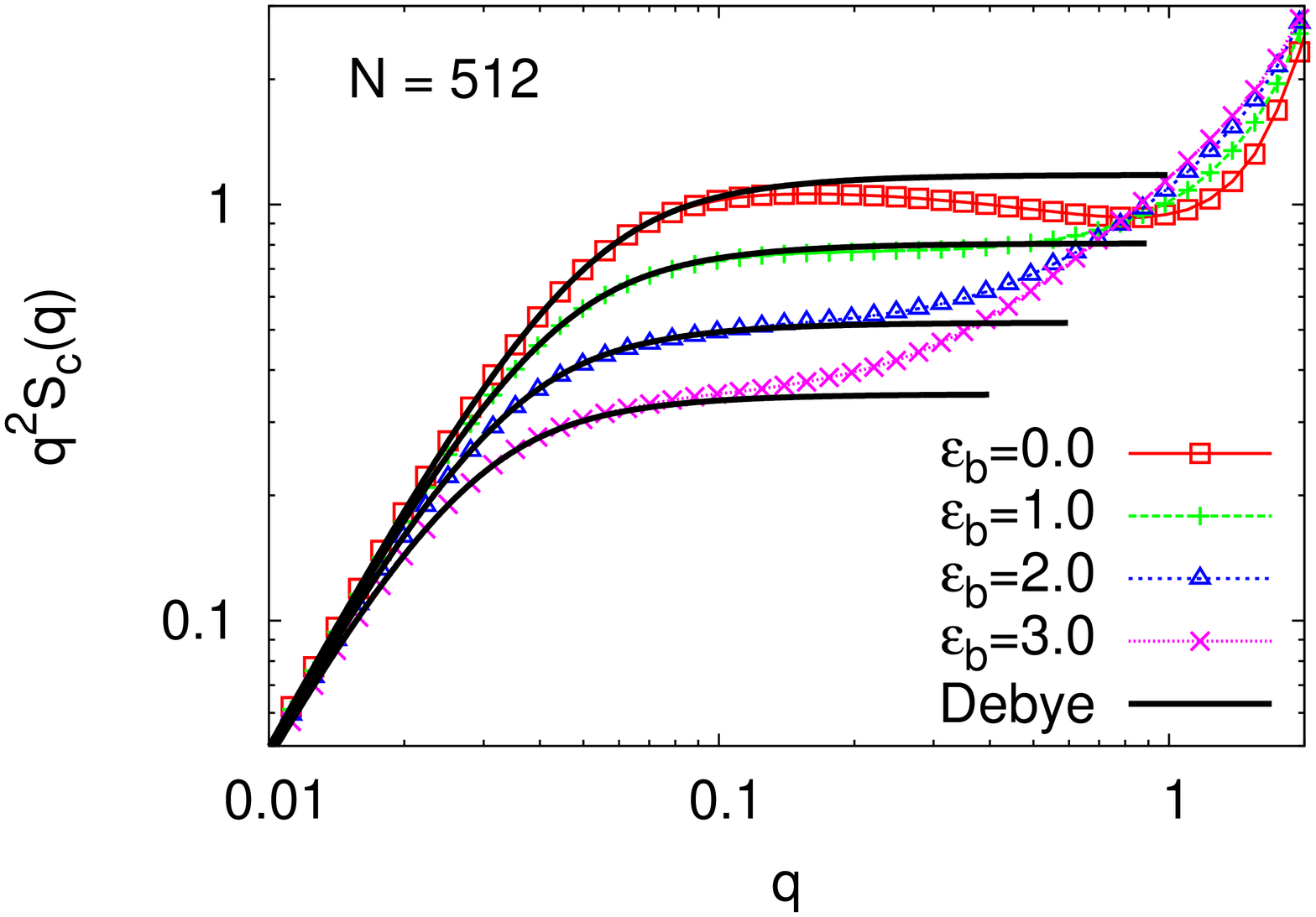}\\
\caption{(a) Structure factors of single chains in a melt, $S_c(q)$,
plotted vs.\ $q$ on log-log scales. (b) Same data as in (a) but in a
Kratky-Plot.
Data are for polymer melts consisting of $n_c=256$ chains of $N=512$ monomers,
and for various values of $\varepsilon_b$, as indicated.
In (a) the theoretical predictions 
$S_c(q)=N(1-q^2 \langle R_g^2 \rangle/3)$ at the Guinier regime for small $q$,
$S_c(q)\sim q^{-2}$ for a Gaussian coil, and $S_c(q) \sim q^{-1}$ for a rigid rod
are shown by solid curves for comparison.
In (b) the Debye function, Eq.~(\ref{Debye}), is also shown for comparison.} 
\label{fig-melt-sqc512}
\end{center}
\end{figure*}

In the following we only focus on polymer melts consisting of 
$n_c=256$ chains of $N=512$ monomers in equilibrium.
Figure~\ref{fig-sq-melt512} shows the results for the scattering
from polymer melts in equilibrium. 
Four different choices
of the stiffness characterized by the bending energy $\varepsilon_b$
are included. 
At small $q$ there is a systematic dependence on $\varepsilon_b$,
reflecting the (rather weak) dependence of the compressibility on
$\varepsilon_b$. It is remarkable that the structure factor of
large values of $q$ is completely independent of the chain stiffness.
As $q$ increases,
the first peak, the so-called amorphous halo for non-crystalline materials,
appears at $q \approx \pi$ measuring the mean inter-particle
distance ($\sim 2$ lattice spacings) in a polymer melt.
The second peak, the so-called Bragg peak, appears at $q \approx 2\pi$
probing the structure on the scale of the monomers ($\sim 1$ lattice spacing).
The scattering from single chains in a melt in equilibrium 
is shown in Fig.~\ref{fig-melt-sqc512}.
In Fig.~\ref{fig-melt-sqc512} one sees that 
$S_c(q) \approx N\exp(-q^2 \langle R_g^2 \rangle/3)
\approx N(1-q^2 \langle R_g^2 \rangle/3)$ for small $q$ in the Guinier regime, then a
cross-over occurs to the power law of Gaussian coils (ideal chains), 
$S(q) \sim q^{-1/\nu_{\rm id}}$ with $\nu_{\rm id}=1/2$. For moderately stiff chains
one observes also a rigid-rod regime 
$S(q) \sim q^{-1}$. The two peaks at $q\approx \pi$ and 
$q\approx 2 \pi$ show up in a similar way as that for the collective structure $S(q)$
in Fig.~\ref{fig-sq-melt512} and are due to the underlying lattice model.
In order to clarify whether single chains in a melt behave as ideal chains
we show the structure factors $S_c(q)$ in a Kratky-plot in Fig.~\ref{fig-melt-sqc512}(b).
The Debye function~\cite{deGennes1979,Cloizeaux1990,Schaefer1999,Higgins1994}
describing the scattering from Gaussian chains,
\begin{equation}
 S_{\rm Debye}(q)=2\frac{\eta-1+\exp(-\eta)}{\eta^2} \quad {\rm with} \quad 
\eta=q^2 \langle R_g^2 \rangle \,,
\label{Debye}
\end{equation} 
is also presented in Fig.~\ref{fig-melt-sqc512}b for comparison.
For fully flexible chains ($\varepsilon_b=0.0$) in a melt 
we see the discrepancy between our
simulation results and the Debye function at the intermediate 
values of $q$, which agrees with the previous
finding~\cite{Wittmer2007i} that polymer chains in a melt are not random
walks. 
While for $\varepsilon_b=0.0$ the deviation from ideality is
clearly recognized due to the minimum in the Kratky-plot near $q=1$,
for $\varepsilon_b \ge 1.0$ there occurs no minimum any longer (for $N=512$).
While the deviations from Gaussian statistics are still obvious from
$\langle \cos \theta(s) \rangle$, Fig.~\ref{fig-hs-melt}, they are not easy 
to extract from $S_c(q)$.
For $\varepsilon_b=1.0$ we see the Kratky plateau over all
intermediate regimes, while the regime decreases as $\varepsilon_b$
increases since the rigid-rod behavior takes over.

\begin{figure*}[t]
\begin{center}
(a)\includegraphics[scale=0.29,angle=270]{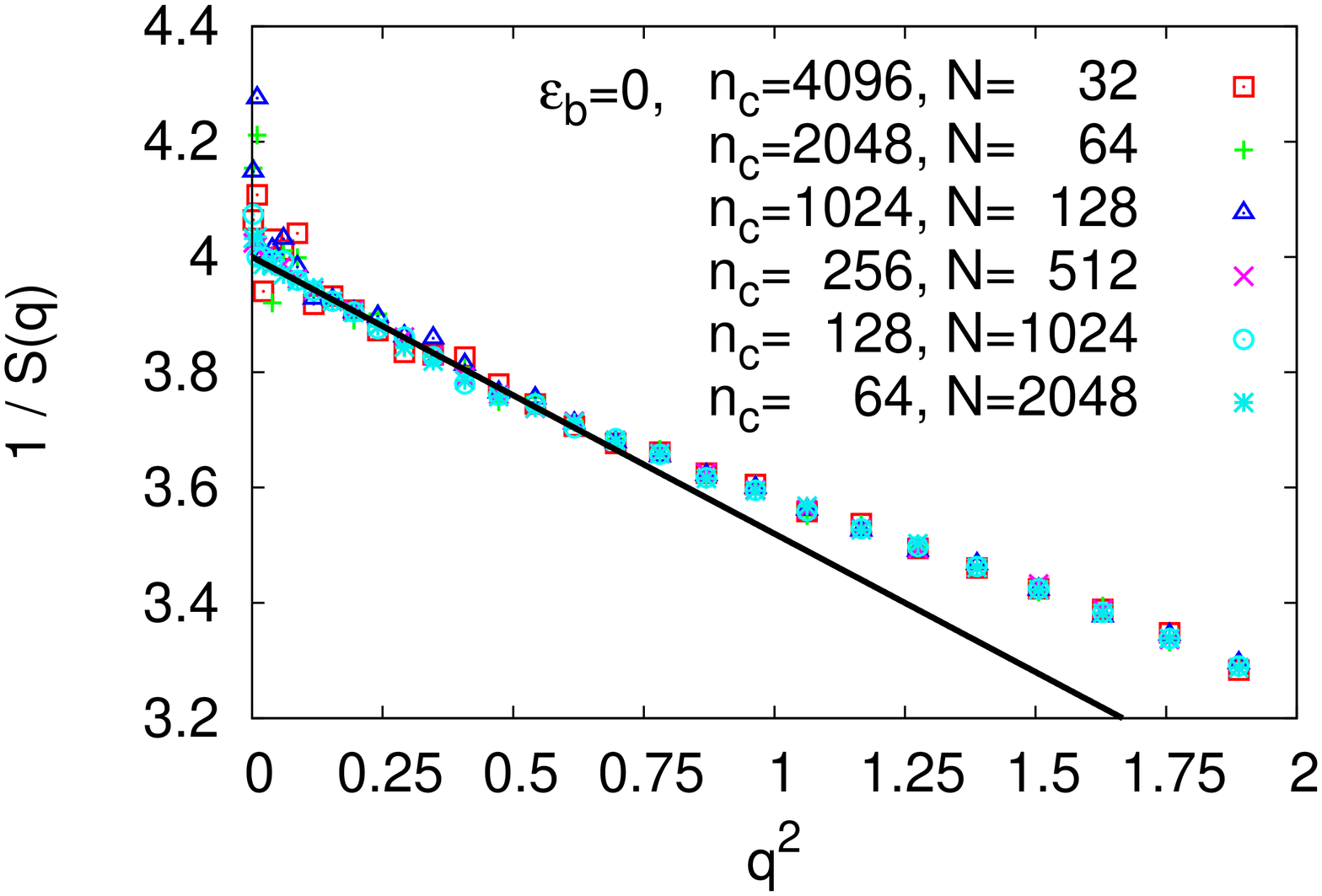} \hspace{0.4cm}
(b)\includegraphics[scale=0.29,angle=270]{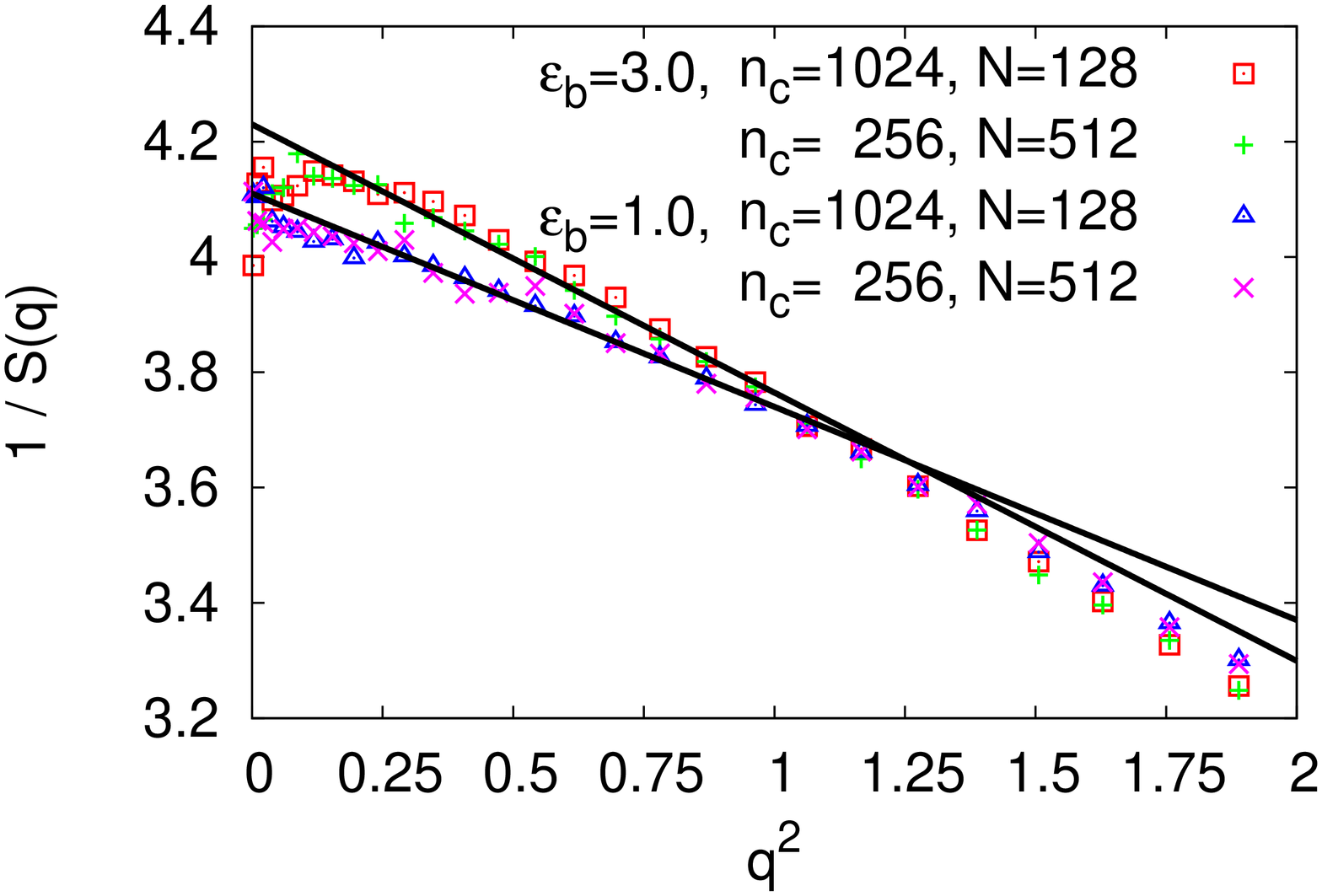}\\
\caption{Inverse of structure factor of polymer melts, $1/S(q)$, plotted vs.\ $q^2$
for $\varepsilon_b=0.0$ (a), and $\varepsilon_b=1.0$, $3.0$ (b).
Several chain lengths $(N-1)$ are chosen, as indicated. The asymptotic behavior of
$1/S(q)$ as $q^2 \approx 0$ is described by straight lines.
Values of the dimensionless compressibility 
$C_g=\left[\lim_{q^2 \rightarrow 0} 1/S(q)\right]^{-1}$
are listed in Table~\ref{table3}.}
\label{fig-kappa}
\end{center}
\end{figure*}

The isothermal compressibility of $\kappa_T$ characterizing the degree
of density fluctuations on large scale is a measure for 
the response of a systems volume to an increase in the system
pressure~\cite{Binder2011,Wittmer2011}, 
$\kappa_T \equiv -(1/V) (\partial V/\partial p)_T$,
and is related to the structure factor $S(q)$ as $q\rightarrow 0$,
\begin{equation}
    \lim_{q \rightarrow 0} S(q)=\rho k_BT \kappa_T =C_g
\label{eq-kappa}
\end{equation}
where $C_g$ is defined as the ``dimensionless compressibility". 
In Fig.~\ref{fig-sq-melt512} we see that as $q \rightarrow 0$
the estimates of the collective structure factor for polymer chains
of size $N=512$ in a melt reach a plateau value 
$\lim_{q \rightarrow 0} S(q) \approx 0.25$. Due to the finite-size effect
and lattice artifact ($q_{\rm min}=2\pi/L$),
we take the same data but plot $1/S(q)$ versus $q^2$ in Fig.~\ref{fig-kappa}.
We see the fluctuations of data near $q^2 \rightarrow 0$.
However, the best fit of the straight line going through our data gives
the estimate of $C_g=\left[ \lim_{q^2 \rightarrow 0} 1/S(q)\right]^{-1}$.  
Results are listed in Table~\ref{table3}. We see that $C_g \in [0.23,0.25]$ and
depends only weakly on the stiffness of the chains.

\begin{figure*}[t]
\begin{center}
(a)\includegraphics[scale=0.29,angle=270]{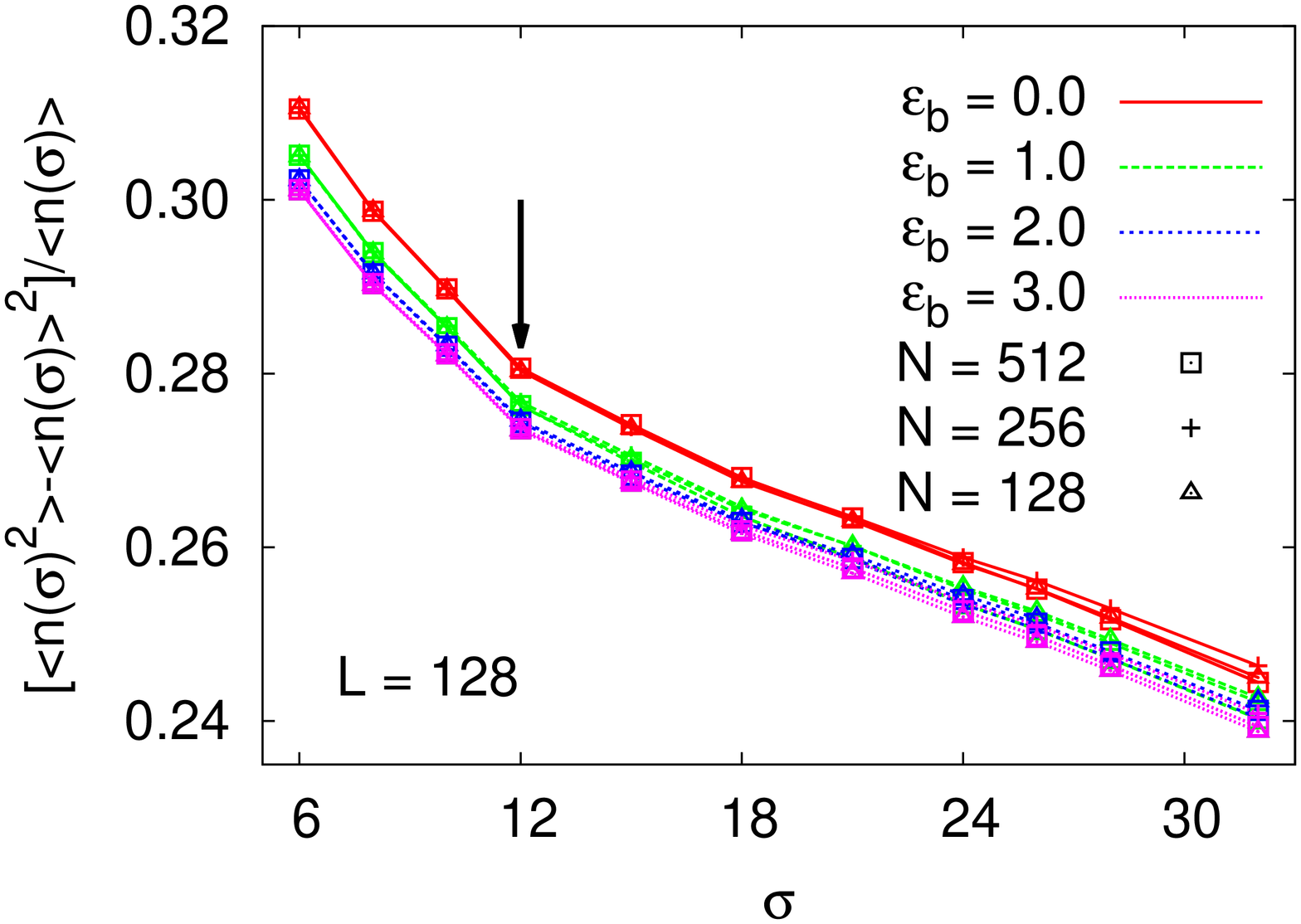} \hspace{0.4cm}
(b)\includegraphics[scale=0.29,angle=270]{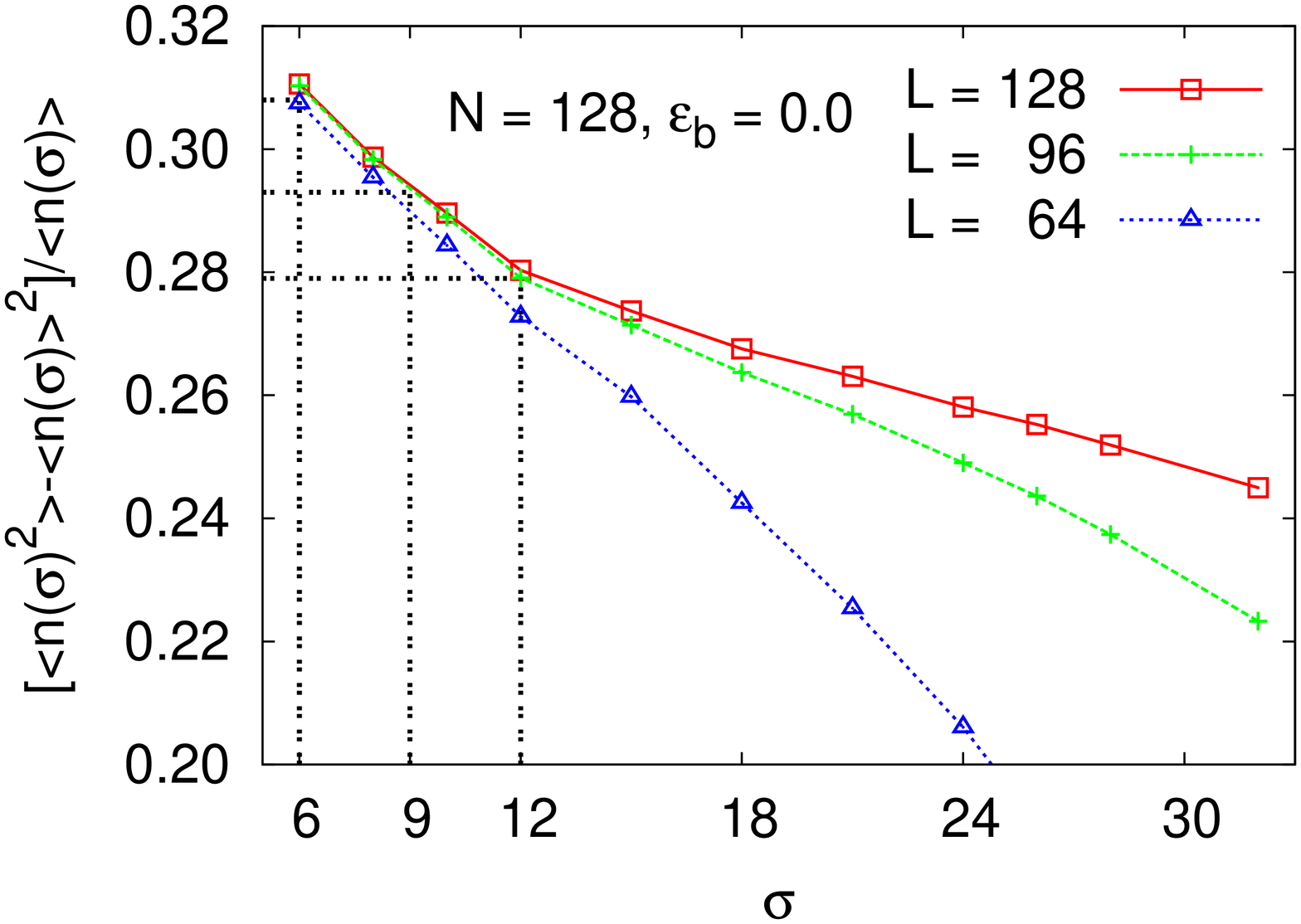}\\
\caption{Local density fluctuations
$\left[\langle n(\sigma)^2 \rangle -\langle n(\sigma) \rangle^2
\right]/\langle n(\sigma) \rangle$ plotted as a function of
$\sigma$ for polymer chains of different stiffnesses in a melt (a) and
for fully flexible chains of fixed chain length $N=128$ in a melt (b).
Three values of chain length $N$ and four values of bending energy $\varepsilon_b$
are chosen, as indicated in (a).
Three different sizes of lattice box are chosen, as indicated in (b).
The locations of $\sigma=\sigma_{\rm min}=(4\pi/3)^{-1/3}(2\pi/L)^{-1}$
for various values of $L$ are indicated by an arrow in (a) and dashed lines in (b).}
\label{fig-cg}
\end{center}
\end{figure*}

In the thermodynamic limit by taking the limit $N \rightarrow \infty$
(but keeping the density $\rho=N/V$ fixed), then taking the limit 
$q \rightarrow 0$, the local density fluctuations in subvolumina 
$V_{\rm sub} \ll V$~\cite{Baschnagel1994, Binder2011}
\begin{equation}
   \frac{\langle n(\sigma)^2 \rangle - \langle n(\sigma) \rangle^2 }
{\langle n(\sigma) \rangle}=k_BT \rho \kappa_T= C_g 
\end{equation}
where $V_{\rm sub}=4\pi \sigma^3/3$.
Choosing $\sigma=12$ such that $q_{\rm min} \approx V_{\rm sub}^{-1/3}$, i.e.
in the limit $q \rightarrow q_{\rm min}=2\pi/L$, the local density fluctuations
estimated from the configurations of polymer melts in equilibrium are
listed in Table~\ref{table1}. The estimates of $C_g \in [0.27,0.28]$ are a
little bit higher than the estimates from the collective structure factor, but
they are compatible. We also check the finite-size effect on the
estimate of the local density fluctuations. Fig.~\ref{fig-cg} shows the local density fluctuations
$[\langle n (\sigma)^2 \rangle - \langle n (\sigma)\rangle^2]/\langle n(\sigma) \rangle$
as a function of $\sigma$. The arrow and dashed lines shown in Fig.~\ref{fig-cg} indicate
the estimate of the compressibility $c_g(L,\varepsilon_b)$  
at $\sigma=\sigma_{\rm min} \approx (4\pi/3)^{-1/3}(2\pi/L)^{-1}$. 
We see that the local density
fluctuations at certain values of $\sigma$ ($\sigma<30$) do not depend on the chain 
lengths $N$ for chains of different stiffnesses,
while the discrepancy between the data sets decreases as the stiffness of chains increases
(see Fig.~\ref{fig-cg}a). For $\sigma<\sigma_{\rm min}$ the local density fluctuations tend to
diverge.
As the size of polymer melts increases ($L$ increases, but $\rho=N/L^3$ is fixed) 
the curves toward to a flat curve for $\sigma>\sigma_{\rm min}$ (see Fig.~\ref{fig-cg}b).
According to the finite-size scaling the estimate of $c_g$ can be obtained by
extrapolating the data of $c_g(L,\varepsilon_b$) 
to $L^{-1} \rightarrow 0$.
  
\section{Conclusions}
\label{conclusion}

We have studied the conformations of polymer chains consisting of fully
flexible and semiflexible chains in a melt based on the bond fluctuation model
at volume fraction $\phi=0.5$.
The initial configurations of polymer
melts are prepared through several procedures. We first generate
NRRWs in a box with periodic boundary conditions in all 
three directions, rearrange NRRWs
by the pre-packing process to reduce the local density fluctuations or
remain NRRWs at the same positions,
and then push off overlapping monomers blocking the same lattice sites. 
We compare the conformations of polymer melts at the end of each process to the
result obtained from the equilibrated configurations.
Applying the additional pre-packing process seems to have no
significant effect on preparing initial configurations of polymer chains in a
melt based on the lattice model after the excluded volume interactions are 
switched on completely.
Namely, the estimates of the mean square 
internal end-to-end distance $\langle R^2(s) \rangle$ and the collective
structure factors $S(q)$ from the configurations generated by the two methods 
(denoted by Pre-packing+EV, and EV in Figs.~\ref{fig-rs-melt512} and
~\ref{fig-sq-melt512i}) are almost the same, and very close to estimates
from the equilibrated configurations. There is also no difference 
of the equilibrating time starting from the initial configurations generated 
by these two methods. 
If, however, one accepts a marginally incomplete
elimination of excluded volume violations ($N_{\rm over} \approx 100$ for
a system of $n_cN=131072$ monomers) one would have significant advantages.
Another possibility of testing this pre-packing process for the
future work would be to use it as a criterion of putting chains 
into the box at the beginning.

In our simulations we combine the algorithms of the local 26 moves, slithering-snake,
and pivot moves (instead of double-bridging moves~\cite{Wittmer2007,Wittmer2011}) for 
equilibrating the system of polymer melts.
Although the total number of monomers $N_{\rm tot}=131072$ which is eight times
smaller than $N_{\rm tot}=1048576$ in Ref.~\cite{Wittmer2007} based on the 
same model for $\varepsilon_b=0$, the estimates
of the mean square end-to-end distance $\langle R_e^2 \rangle$, 
the mean square gyration radius $\langle R_g^2 \rangle$, the mean square bond
lengths $\langle \vec{b}^2 \rangle$, and the dimensionless compressibility
$C_g$ are all in perfect agreement with those results given in Ref.~\cite{Wittmer2007}
at fixed chain size $N$. For fully flexible and moderately stiff chains the ratio 
$\langle R_e^2 \rangle/\langle R_g^2 \rangle \approx 6$ as expected for ideal chains.
For moderately stiff chains in a melt the internal mean square end-to-end distance 
$\langle R^2(s) \rangle$ is well described by the freely rotating chain model.
Results of the probability distributions of reduced end-to-end distance
$r_e=(R_e^2/\langle R_e^2 \rangle)^{1/2}$ and reduced gyration radius
$r_g=(R_g^2/\langle R_g^2 \rangle)^{1/2}$ for polymer chains in a melt
for various values of $N$ and $\varepsilon_b$ 
show the nice data collapse, and are described by universal functions,
Eqs.~(\ref{eq-pre}) and (\ref{eq-prg}), for ideal chains.
The collective structure factors $S(q)$ for the whole polymer melts and the 
standard structure factor $S_c(q)$ for single chains in a melt are also
calculated and compared with the theoretical predictions.
A detailed investigation of $S_c(q)$ in a Kratky-plot for $\varepsilon_b=0.0$,
$1.0$, $2.0$, and $3.0$ shows that for fully flexible chains in a melt there
is a significant deviation from the Debye function for Gaussian chains at
the intermediate values of $q$ 
as found by Wittmer et al.~\cite{Wittmer2007i}, 
while for $\varepsilon_b=1.0$ it
is perfectly described by the Debye function. 
Since real polymers (such as polystyrene) exhibit some local
chain stiffness, it is clear that the deviations from Gaussian statistics
found for fully flexible chains in melts in the Kratky-plot will be
very difficult to test experimentally.
{However, since only chains up to $N=512$ are considered
in our simulations, such a deviation may also occur as the chain length
increases. Careful investigation of the structure factor 
and the mean square internal end-to-end distance
for much longer semiflexible polymer chains in a melt will be required
to clarify this matter.}
The dimensionless compressibility $C_g$ determined by the estimates of
the collective structure factors $S(q)$ for small $q$ 
and the local density fluctuations
$(\langle n^2 \rangle - \langle n \rangle^2)/\langle n \rangle$
within a sphere of radius $\sigma$
are compatible from our simulations for all cases.
We hope that the present work will help to the further development of
a coarse-graining approach by using the BFM as an underlying microscopic
model and will be useful for the interpretation
of corresponding experiments searching for excluded volume effect
in the scattering function of polymers in melts.

\section{Acknowledgments}

  I am indebted to K. Binder and K. Kremer for stimulating discussions
and for carefully reading the manuscript, and also thank 
L. Moreira and W. Paul for their helpful discussions. 
I thank the Max Planck Institute for Polymer Research for the
hospitality while this research was carried out.
I also thank the John von Neumann Institute for Computing
(NIC J\"ulich) for a generous grant of computer time
and the Rechenzentrum Garching (RZG), the supercomputer center of
the Max Planck Society, for the use of their computers.


\begin{thebibliography}{99}

\bibitem{Flory1969} P. J. Flory, {\it Statistical mechanics of chain molecules},
Wiley press, New York (1979).

\bibitem{deGennes1979} P. G. de Gennes, {\it Scaling Concepts in 
polymer physics}, Cornell University Press: Itharca, New York. (1979).

\bibitem{Binder1995} K. Binder (ed.), {\it Monte Carlo and molecular dynamics
simulations in polymer science}, Oxford University Press, New York (1995).

\bibitem{Kotelyanskii2004} M. Kotelyanskii and D. N. Theodorou (ed),
{\it Simulation methods for polymers}, Marcel Dekker, New York (2004).

\bibitem{Binder2008} K. Binder and W. Paul, 
Macromolecules {\bf 41}, 4537 (2008).

\bibitem{Murat1998} M. Murat and K. Kremer, 
J. Chem. Phys. {\bf 108}, 4340 (1998).

\bibitem{Plathe2002} F. Müller-Plathe, 
ChemPhysChem {\bf 3}, 754 (2002).

\bibitem{Harman2007} V. A. Harmandaris, D. Reith, N. F. A. van der Vegt, and K. Kremer, 
Macromol.  Chem. Phys. {\bf 208}, 2109 (2007).

\bibitem{Harman2006} V. A. Harmandaris, N. P. Adhikari, N. F. A. van der Vegt, and K. Kremer, 
Macromolecules {\bf 39}, 6708 (2006).

\bibitem{Gujrati2010} P. D. Gujrati and A. I. Leonov (Ed.), 
{\it Modeling and simulations in polymers}, Wiley (2010).

\bibitem{Vettorel2010} T. Vettorel, G. Besold, and K. Kremer, 
Soft Matter {\bf 6}, 2282 (2010).

\bibitem{Zhang2013} G. Zhang, K. Ch. Daoulas, and K. Kremer, 
Macromol. Chem. Phys. {\bf 214}, 214 (2013).

\bibitem{Zhang2014} G. Zhang, L. A. Moreira, T. Stuehn, K. Ch. Daoulas, and K. Kremer,
ACS Macro Lett. {\bf 3} 198, (2014).

\bibitem{Carmesin1988} I. Carmesin and K. Kremer, 
Macromolecules {\bf 21}, 2819 (1988).

\bibitem{Wittmann1990} H.-P. Wittmann and K. Kremer, Comp. Phys. Commun.
{\bf 61}, 309 (1990).

\bibitem{Deutsch1991} H. P. Deutsch and K. Binder, 
J. Chem. Phys. {\bf 94} 2294 (1991).

\bibitem{Paul1991} W. Paul, K. Binder, D. W. Heermann, and K. Kremer, 
J. Phys. II {\bf 1} 37 (1991).

\bibitem{Wittmer2011} J. P. Wittmer, A. Cavallo, H. Xu, J. E. Zabel,
P. Poli\'nska, N. Schulmann, H. Meyer, J. Farago, A. Johner, S. P. Obukhov,
and J. Baschnagel, J. Stat. Phys. {\bf 145}, 1017 (2011).

\bibitem{Yamakawa1971} H. Yamakawa, {\it Modern theory of polymer solutions},
Harper and Row, New York (1971).

\bibitem{Wittmer2007i} J. P. Wittmer, P. Beckrich, A. Johner, A. N. Semenov,
S. P. Obukhov, H. Mayer, and J. Baschnagel, EPL {\bf 77} 56003 (2007).

\bibitem{Auhl2003} R. Auhl, R. Everaers, G. S. Grest, K. Kremer, and S. J. Plimpton,
J. Chem. Phys. {\bf 119}, 12718 (2003).

\bibitem{Moreira2014} L. A. Moreira, F. M\"uller, T. St\"uhn, and K, Kremer,
unpublished (2014).

\bibitem{Hsu2014} H.-P. Hsu, J. Chem. Phys., {\bf 141}, 164903 (2014).

\bibitem{Wittmer1992} J. P. Wittmer, W. Paul, K. Binder, Macromolecules {\bf 25},
7211 (1992).

\bibitem{Wittmer2007} J. P. Wittmer, P. Beckrich, H. Mayer, A. Cavallo, A. Johner,
and J. Baschnagel, J. Phys. Rev. E {\bf 76}, 011803 (2007).

\bibitem{Grosberg1994} A. Yu. Grosberg, and A. R. Khokhlov, 
{\it Statistical Physics of Macromolecules} AIP Press, NY, (1994).

\bibitem{Rubinstein2003} M. Rubinstein and R. H. Colby, {\it Polymer Physics},
Oxford University Press, Oxford, (2003).

\bibitem{Hsu2010} H.-P. Hsu, W. Paul, and K. Binder, Macromolecules {\bf 43},
3094 (2010).

\bibitem{Yamakawa1970} H. Yamakawa, {\it Modern theory of polymer solutions},
Clarendon, Oxford (1970).

\bibitem{Fujita1970} H. Fujita and T. Norisuye, J. Chem. Phys. {\bf 52}, 1115 (1970).

\bibitem{Denton2002} A. R. Denton and M. Schmidt, J. Phys.: Condens. Matter
{\bf 14}, 12051 (2002).

\bibitem{Froehlich2013} M. G. Fr\"ohlich and T. D. Sewell, 
Macromol. Theory Simul. {\bf 22}, 344 (2013).

\bibitem{Lhuillier1988} D. Lhuillier,
J. Phys. France {\bf 49}, 705 (1988).

\bibitem{Cloizeaux1975} J. des Cloizeaux, J. Phys. France {\bf 36}, 281 (1975).

\bibitem{Fisher1966} M. E. Fisher, J. Chem. Phys. {\bf 44}, 616 (1966).

\bibitem{Cloizeaux1990} J. Des Cloizeaux and G. Jannink, 
{\it Polymers in Solution: Their Modeling and Structure}, Clarendon, Oxford (1990).

\bibitem{Schaefer1999} L. Sch\"afer, {\it Excluded Volume Effects in Polymer 
Solutions as Explained by the Renormalization Group}, Springer, Berlin, (1999).

\bibitem{Higgins1994} J. S. Higgins and H. C. Benoit, 
{\it Polymers and Neutron Scattering}, Clarendon, Oxford (1994).

\bibitem{Baschnagel1994} J. Baschnagel and K. Binder,
Physica A {\bf 204}, 47 (1994).

\bibitem{Binder2011} K. Binder and W. Kob, {\it Glassy materials and disordered solids:An
introduction to their statistical mechanis}, 2nd ed., World Scientific press, Singapore (2011).

\end{thebibliography}
\end{document}